\documentclass[twocolumn]{aastex631}
\shorttitle{The IMFs of young clusters in NGC 1313}
\shortauthors{Koo et al.}
\graphicspath{{./}{figures/}}
\usepackage{hyperref}
\begin{document}

\title{Initial Mass Functions of Young Stellar Clusters from the Gemini Spectroscopic Survey of Nearby Galaxies. II. Young Clusters in NGC 1313}

\correspondingauthor{Beomdu Lim}
\email{blim@kongju.ac.kr}

\author[0000-0001-8969-0009]{Jae-Rim Koo}
\affiliation{Earth Environment Research Center, Kongju National University, 56 Gongjudaehak-ro, Gongju-si, Chungcheongnam-do 32588, Republic of Korea}
\affiliation{Korea Astronomy and Space Science Institute, 776 
Daedeok-daero, Yuseong-gu, Daejeon 34055, Republic of Korea}

\author[0000-0001-9263-3275]{Hyun-Jeong Kim}
\affiliation{Korea Astronomy and Space Science Institute, 776 
Daedeok-daero, Yuseong-gu, Daejeon 34055, Republic of Korea}

\author[0000-0001-5797-9828]{Beomdu Lim}
\affiliation{Earth Environment Research Center, Kongju National University, 56 Gongjudaehak-ro, Gongju-si, Chungcheongnam-do 32588, Republic of Korea}
\affiliation{Korea Astronomy and Space Science Institute, 776 
Daedeok-daero, Yuseong-gu, Daejeon 34055, Republic of Korea}
\affiliation{Department of Earth Science Education, Kongju National University, 
56 Gongjudaehak-ro, Gongju-si, Chungcheongnam-do 32588, Republic of Korea}



\begin{abstract}
We present a spectroscopic study of young stellar clusters 
in the barred spiral galaxy NGC 1313. Integrated light spectra 
of 11 clusters, obtained using the GMOS-S instrument 
on the 8.1 m Gemini South telescope, are analyzed using 
a simple stellar population model. A subsolar metallicity 
($Z = 0.008$) is adopted, consistent with previous studies. 
Cluster ages are constrained primarily through absorption 
lines and prominent emission bands of Wolf–Rayet stars. 
Utilizing these constraints, we match the observed 
spectra with synthetic counterparts generated from the simple 
stellar population model, determining key physical parameters 
including age, cluster mass, and the underlying 
initial mass function (IMF). Furthermore, the impact of stochastic 
sampling on the derived parameters of several low-mass clusters 
is rigorously evaluated using Monte Carlo simulations. The sampled 
clusters exhibit ages ranging from 2.5 to 300 Myr and stellar masses 
between $2.8 \times 10^3 \ M_{\odot}$ and 
$2.6 \times 10^5 \ M_{\odot}$. Notably, for stellar masses exceeding 
$0.8 M_{\odot}$, the power-law index ($\Gamma$) of 
the underlying IMFs is found to be smaller than the 
standard Salpeter/Kroupa IMF. Furthermore, a correlation is 
observed where more massive clusters tend to possess top-light 
IMFs. This finding aligns with trends observed in 
the young clusters of the Antennae Galaxies, despite 
the differing mass scales between the two systems. 
Our results suggest that applying a universal standard 
IMF to spatially unresolved systems warrants caution, 
given the inherent complexities revealed in this study.
\end{abstract}



\section{Introduction} \label{sec:1}
The stellar initial mass function (IMF) is the mass 
distribution of stars with the same origin 
\citep{1955ApJ...121..161S}. The hypothesis that 
the masses of stars are drawn from the same underlying 
distribution of stellar mass has been accepted and 
examined \citep{1979ApJS...41..513M,2001MNRAS.322..231K,
2010ARA&A..48..339B}. Meanwhile, a variation of the IMF 
in the high-mass regime has also been suggested for 
young massive clusters in the Galaxy 
\citep{2004AJ....127.1014S,2008ApJ...675.1319H,
2013AJ....145...46L,2019ApJ...870...44H}. Low-to-moderate 
redshift galaxies with high star formation rates (SFRs) 
tend to have shallow IMFs \citep{2011MNRAS.415.1647G}. 
The universality of the IMF has long been a controversial 
issue in various fields of astronomy.

Young stellar clusters are the ideal objects 
to examine the universality or the diversity 
of the IMF. Most stars form in stellar clusters or 
associations \citep{2003ARA&A..41...57L,
2003AJ....126.1916P}. This fact allows us 
to assume an instantaneous star formation, 
i.e., a simple star formation history. Stars 
form over a wider range of stellar masses as 
clusters are more massive \citep{2000ApJ...539..342E,
2004MNRAS.348..187W,2017IAUS..316..357L}. In 
addition, the most massive stars with very short 
lifetime are still observable in the extremely 
young clusters. 

The metallicity and density of molecular 
clouds are the factors that can affect 
the Jeans mass and thereby the IMF 
\citep{2012MNRAS.422.2246M}. In the Galaxy, 
the metallicity of open clusters decreases 
with the distance from the Galactic center 
\citep{2012AJ....144...95Y,2025AJ....169..214Y}. 
The surface densities of hydrogen molecules and 
atoms also decrease toward the outer Galaxy 
\citep{1984ApJ...276..182S,1990A&A...230...21W}. 
However, since the majority of observable 
open clusters are concentrated within 3 kpc from 
the Sun \citep{2018A&A...618A..93C}, the variations 
in metallicity and cloud density is not large enough 
to determine how environmental influence affect the 
star formation process. Furthermore, young massive 
clusters are very rare and 
are severely obscured \citep[etc.]{2010A&A...524A..82L,
2012MNRAS.419.1871D,2013AJ....145...46L,2015MNRAS.446.3797H,
2019ApJ...870...44H}.

We have initiated a systematic survey of 
young stellar clusters in nearby galaxies. 
The target galaxies have different star-forming 
environments in terms of metallicity, star formation 
rate (SFR), and the external forces of satellite 
galaxies. The first target galaxies of this survey 
were the starburst galaxies NGC 4038/9 
\citep{2025AJ....169....7K}. We analyzed the integrated 
light spectra of seven young massive clusters 
in the galaxies by means of a simple stellar 
population model. Interestingly, there is a 
relationship between the IMF and cluster mass. 
More massive clusters appear to be dominated 
by low-mass star population, i.e., a top-light IMF. 
The IMF of the most massive cluster ID 11 has a power-law index 
$\Gamma = -2.0^{+0.3}_{-0.0}$, which is consistent 
with the result of a previous study \citep{2010ApJ...710.1746G} 
when adopting a single burst model.

This relationship showed a dependency of the IMF on 
the cluster mass, but does not imply how 
star-forming environment has influence on 
the IMF; we need further investigations 
of young clusters in other galaxies. In this paper, 
our contribution extends to the metal-poor galaxy 
NGC 1313 which has two strong spiral arms with 
a bar structure. This galaxy is 4.2 -- 4.6 Mpc away from the Sun \citep{2009AJ....138..332J,2015ApJ...799...19Q,2018ApJS..235...23S}. 
This galaxy contains the total stellar mass of 
$2.6\times 10^9 \ M_{\sun}$, and stars are forming 
at a rate of 1.15 $M_{\sun}$ yr$^{-1}$ 
\citep{2015AJ....149...51C}. The observation of 
the atomic hydrogen line revealed an ongoing tidal 
interaction between NGC 1313 and its satellite 
galaxy \citep{1994MNRAS.269.1025P}. The recent 
increase of SFR and star 
formation efficiency in the south 
arm are likely related to this tidal interaction 
\citep{2007IAUS..241..435L,2012MNRAS.423..213S,
2013A&A...554A...8S}.

NGC 1313 hosts low-metallicity environments. 
\citet{2017A&A...603A.119H} measured the metallicity 
of the young cluster NGC 1313-379 to be [Fe/H] $= -0.84 \pm 0.07$, which is 
lower than stellar clusters in the Large Magellanic Cloud. Their 
subsequent study of the chemical abundance of five additional 
clusters confirmed this subsolar metallicity \citep{2022AJ....164...89H}. 
Furthermore, a photometric study inferred $Z = 0.008$ by 
comparing the spectral energy distributions of young clusters 
with synthetic models \citep{2021ApJ...909..121M}.

The Legacy ExtraGalactic UV Survey (LEGUS) is a $Hubble$ treasury 
program designed to provide a homogeneous imaging data set of 
nearby galaxies in five bands from the UV to the NIR 
\citep[LEGUS;][]{2015AJ....149...51C}\footnote{\url{https://archive.stsci.edu/prepds/legus/dataproducts-public.html}}. 
A large number of stellar clusters in NGC 1313 were identified 
through this imaging survey (673 clusters younger than 300 Myr -- 
\citealt{2017ApJ...841..131A}). Later, young, highly embedded 
clusters were additionally discovered from infrared images 
\citep{2021ApJ...909..121M}. 

There are local differences in the properties of stellar 
clusters and molecular clouds within NGC 1313 
\citep{2024ApJ...964...13F}. The total stellar masses of 
individual clusters appear to be larger in the interarm 
regions, but many of them are old. The majority of young 
clusters are distributed along the spiral arms, particularly 
the north arm. The spiral arms not only have a larger mass 
of molecular clouds, but they also have higher kinetic energy 
and surface density than clouds in the regions between 
the spiral arms.

The properties of young clusters and molecular clouds 
were often compared with those of the flocculent spiral galaxy 
NGC 7793, owing to similarities in their total stellar 
mass, metallicity, star formation rate, and morphology \citep{1997MNRAS.288..726W,2015AJ....149...51C,2015ApJ...812...39S}. 
NGC 1313 contains a larger number of young massive clusters than NGC 7793 
\citep{2015AJ....149...51C,2024ApJ...964...12F}. Compared to the latter, 
the molecular clouds in NGC 1313 has higher kinetic energy, and a larger 
fraction of the clouds is in a near virial equilibrium 
\citep{2024ApJ...964...12F}. 

The environment of NGC 1313 presents a stark contrast to 
the star-forming conditions previously observed in NGC 4038/9, 
making it an ideal candidate for a comparative analysis 
of the stellar IMF. Furthermore, this study aims to produce 
results that enable a robust comparison with NGC 7793, a galaxy 
characterized by a similar star-formation environment. Such 
comparisons are expected to provide a more comprehensive 
understanding of IMF variations across diverse galactic 
contexts.

The remainder of this paper is organized as follows: 
Section~\ref{sec:2} describes the spectroscopic observations 
and data reduction procedures. Section~\ref{sec:3} presents the 
primary results of our analysis, followed by a discussion of 
influencing factors in Section~\ref{sec:4}. Finally, we summarize 
our key findings and conclusions in Section~\ref{sec:5}.

\begin{figure}[t]
\epsscale{1.0}
\plotone{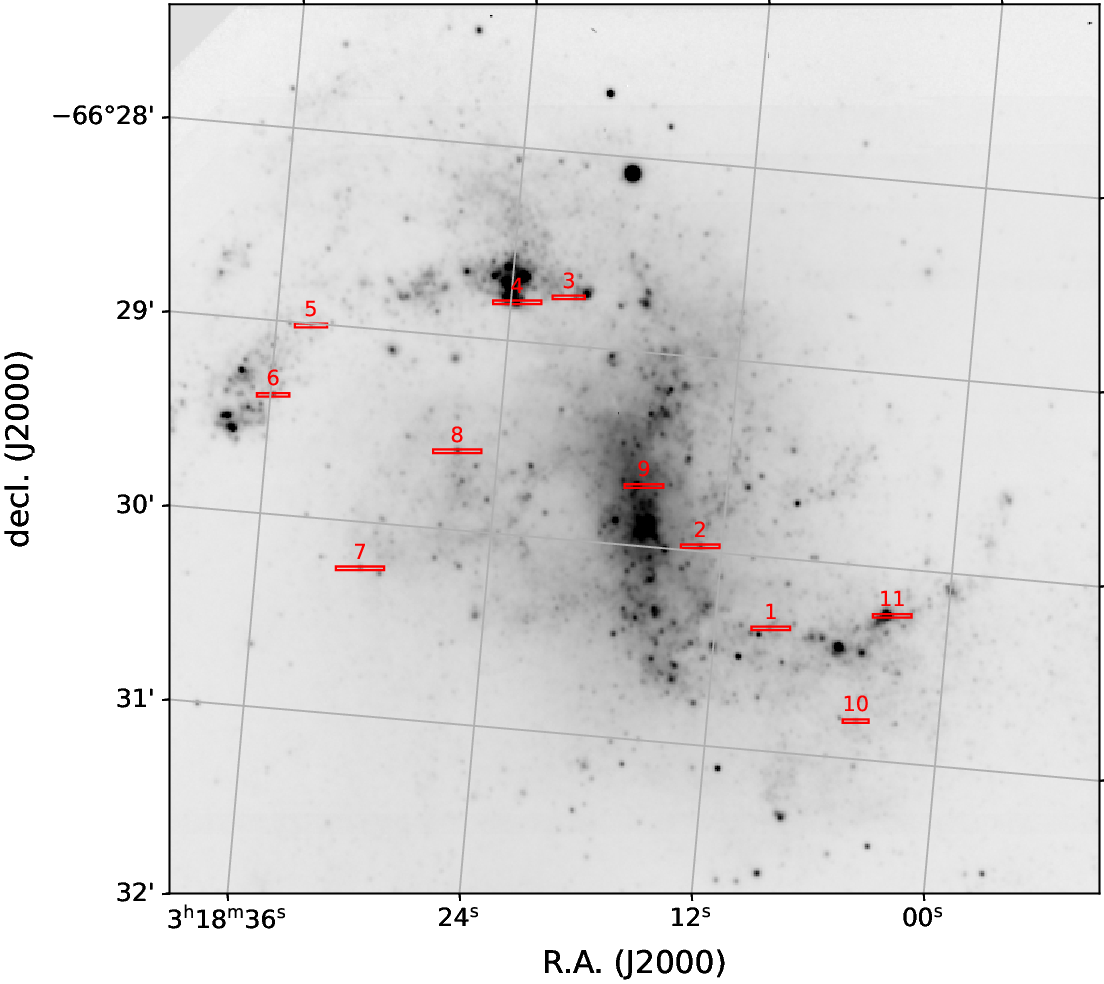}
\caption{GMOS-S image of NGC~1313. The slit positions are shown by 
rectangular boxes with the cluster IDs.}\label{fig1}
\end{figure}

\begin{figure*}[t]
\epsscale{1.0}
\plotone{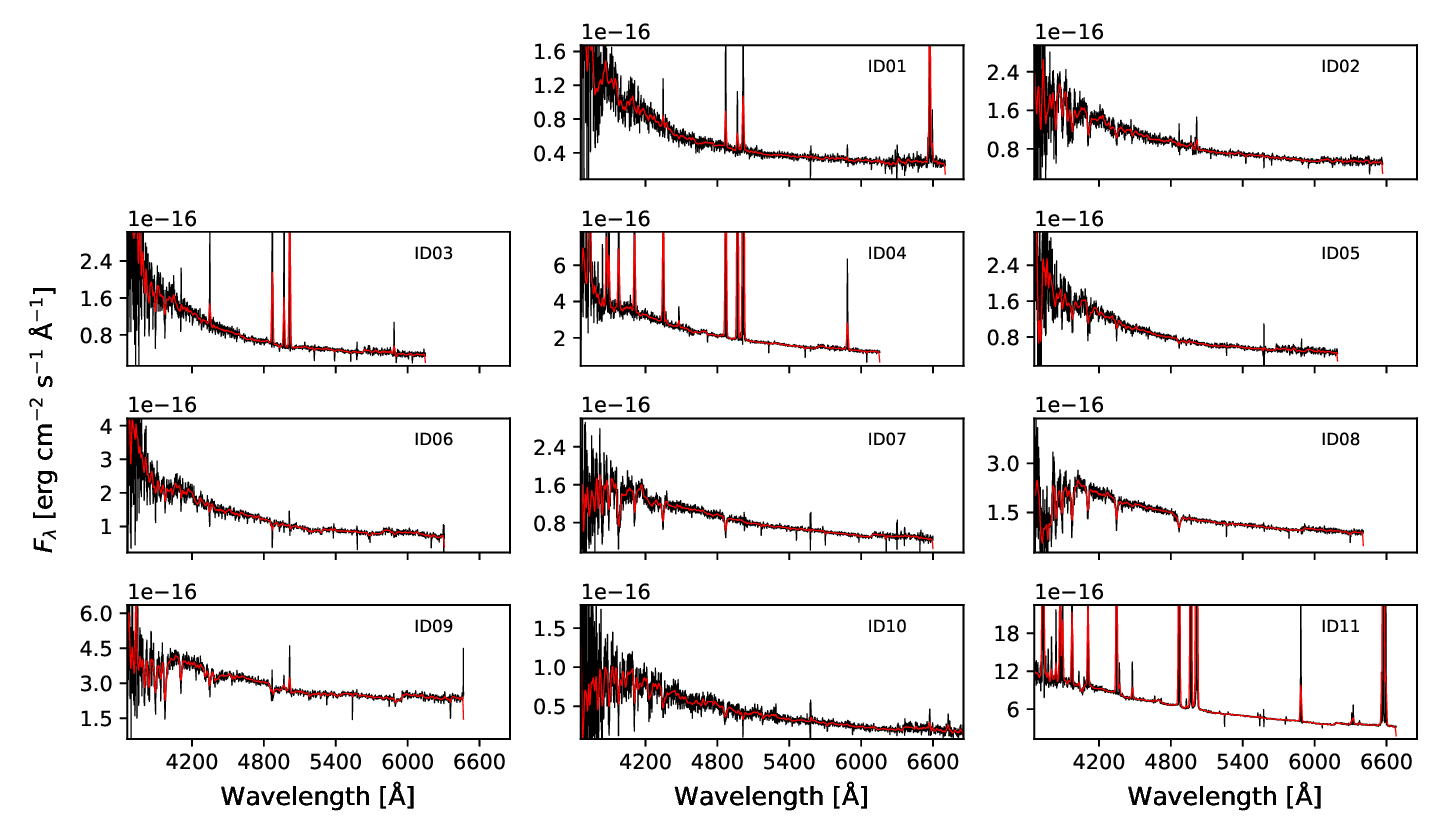}
\caption{Flux-calibrated spectra of 11 clusters. The IDs of 
individual clusters are labeled in the upper right corner of 
each panel. The red curves represent the smoothed spectra 
to highlight spectral features.
}\label{fig2}
\end{figure*}

\section{Data} \label{sec:2}
\subsection{Observation}
LEGUS published the catalogs of cluster candidates spread 
over two $Hubble$ fields in NGC~1313 \citep{2017ApJ...841..131A}. 
These catalogs contain photometric data in five optical 
passbands and physical parameters of clusters, such as 
ages, masses, and reddening, inferred from analysis 
of their spectral energy distributions. We merged the 
catalogs into a master catalog after removing duplicates. 
A total of 77 clusters brighter than 20 mag in the F555W 
band (equivalent to $V$ band) were included in our target 
list. Note that the physical parameters of the selected 
clusters were derived with high-confidence levels in 
the previous study.

We performed $g$-band pre-imaging observations to design and 
fabricate a mask for the Gemini Multi-Object Spectrograph 
(GMOS) at Gemini South (Program ID: GS-2023B-Q-116). 
Multi-object spectroscopic observations of 11 clusters 
from our target list were conducted using a $0\farcs75$ slit 
width and the B600 grating. The grating was centered 
at 4000, 4500, and 5000 \AA\ to bridge the physical CCD chip 
gaps and ensure broad spectral coverage for clusters distributed 
across the GMOS mask. A total of 17 frames were obtained over three nights 
from 2024 October 25 to November 5, with an exposure 
time of 900 s per frame. 

In addition, the standard star EG21 (DA) was observed using the GMOS 
in a long-slit mode with a slit width of (0$\farcs75$) on the same 
nights as the observations of the clusters. To account for 
wavelength shifts caused by slit offsets along the dispersion 
axis, we performed wavelength dithering for the science targets. 
Consequently, the standard star was observed over a broader spectral range 
to ensure a robust response curve across the entire combined 
spectral coverage of the MOS mask. The B600 
grating was thus centered at 4000, 5000, and 6000 \AA. 

The observational data of the clusters were reduced 
in a standard manner using the Gemini IRAF package. The wavelength 
solution derived from arc spectra was applied to the two-dimensional 
spectra of the clusters. The one-dimensional (1D) spectra of 
individual clusters were then extracted using the custom Python code, 
as described in our previous study \citep{2025AJ....169....7K}.

Prior to extraction, we explicitly accounted for the 
wavelength-dependent instrumental response, including the quantum 
efficiency (QE) variations across the CCD chips\footnote{The QE curves 
of the GMOS detector were adopted from Data Reduction for Astronomy 
from Gemini Observatory North and South (DRAGONS).}. This step was 
essential to ensure that the detector-level characteristics were 
homogenized before the final flux calibration, especially as we 
transitioned from the IRAF-based pre-processing to our custom 
extraction routine. The aperture radii of 10 or 15 pixels, equivalent 
to $1\farcs6$ or $2\farcs5$, were used to minimize contamination 
from neighboring sources. The regions outside the adopted apertures 
along given slits were considered as sky background. The sky 
background signals were subtracted from the 1D spectra, followed 
by detrending if required. Finally, the mean atmospheric extinction 
at Gemini South \citep{1983MNRAS.204..347S} was applied to the individual 
spectra.

We selected an observing set of the standard 
star observed on 2024 November 5. The long-slit 
spectra of the standard star were reduced using 
DRAGONS v3.2.0 \citep[Data Reduction for Astronomy from 
the Gemini Observatory North and South;][]{2023RNAAS...7..214L,2024zndo..10841622S}. 
The correction for the quantum efficiency 
is included in the reduction with DRAGONS. The spatial 
profile of signals along the slit at a given 
wavelength was fit by a Gaussian distribution. 
The 1D spectra were then extracted within an aperture 
radius of $5\,\sigma$ from the best-fit Gaussian. 
Subsequently, the individual spectra of this star 
were corrected for the atmospheric extinction as above. 

The GMOS does not contain atmospheric dispersion 
corrector in its optical components. Therefore, 
the atmospheric refraction results in different 
levels of slit-losses as the position angles of 
slits in the mask deviate from the parallactic 
angle with time \citep{1982PASP...94..715F}. 
This effect highly depends on 
the seeing size relative to the slit width 
($0\farcs75$). We corrected the slit-losses 
for all 1D spectra using \texttt{spec.lightloss2} 
function from \texttt{Ian's Astro-Python Code}\footnote{\url{https://crossfield.ku.edu/python/}}.
To address the potential uncertainties associated with slit loss, we assumed that the telescope was accurately positioned and tracked to keep the targets centered within the slit mask during the observations. Moreover, the point spread functions of our target clusters were treated as symmetric 2D-Gaussian profiles. Although these young clusters are extended systems rather than single point sources, a symmetric 2D-Gaussian approximation serves as a reasonable first-order baseline for our slit loss correction. Observationally, the seeing during our runs ranged from $0\farcs9$ to $1\farcs0$ (FWHM), slightly exceeding our slit width of $0\farcs75$. Crucially, while tighter seeing conditions can cause the slit loss fraction to fluctuate drastically with minute centering offsets, a seeing size comparable to or slightly larger than the slit width yields a flatter, more predictable light fraction curve, making the correction less sensitive to centering jitters. Furthermore, our observations were conducted at relatively low airmass conditions with a slit position angle of approximately $-85^\circ$, which effectively minimized differential atmospheric refraction and ensured that slit losses remained nearly uniform and achromatic across the entire observed wavelength range.

The spectra of the clusters observed under 
the best seeing condition were adopted as 
references. The other spectra were then scaled 
to the flux levels of the references. The final 
spectrum of a given cluster was obtained from 
the median combined spectra. Also, the spectra of 
the standard star were combined into a single 
spectrum. The full response curve was derived 
from comparison of the flux-calibrated spectra 
with the observed spectra of the standard star. 
Finally, the final spectra of individual clusters 
were calibrated in flux using the full response 
curve. Figure~\ref{fig2} displays the calibrated 
spectra of 11 clusters.

\begin{figure}[t]
\epsscale{1.0}
\plotone{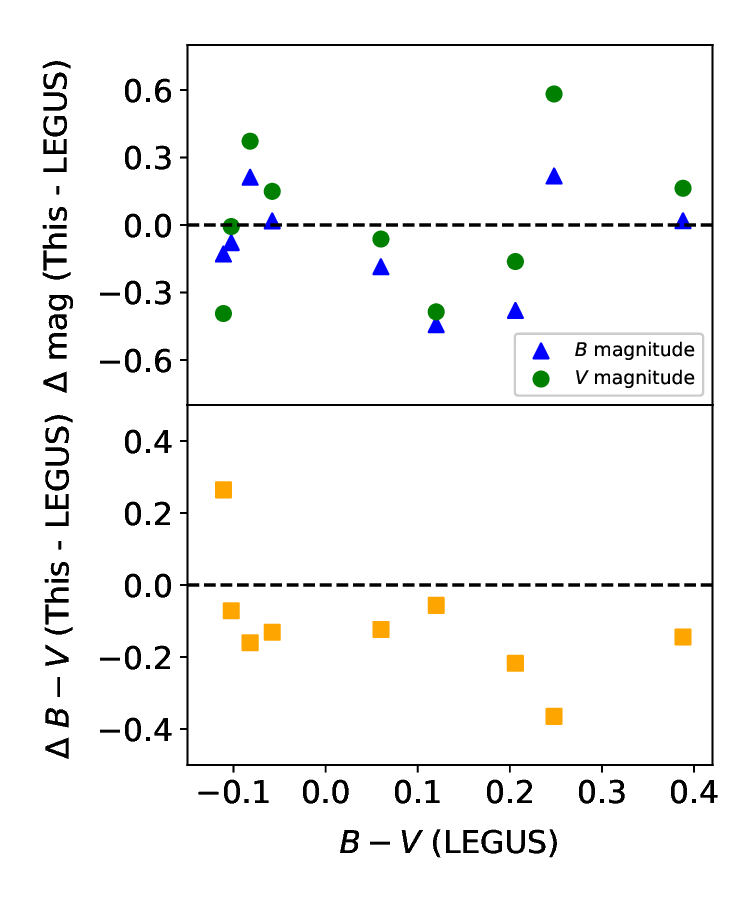}
\caption{Comparison between our synthetic photometry 
(from calibrated spectra) and LEGUS photometry. The blue 
and green dots represent photometric differences in the $B$ 
and $V$ bands, respectively.}\label{fig3}
\end{figure}

We performed synthetic photometry on the final spectra 
using \texttt{pyphot} \citep{fouesneau2025pyphot} 
and compared with the LEGUS catalog, with exception 
for a blended cluster ID 11. Figure~\ref{fig3} presents 
a comparison of the two band magnitudes. The mean 
and standard deviation of the differences are
$\Delta B= -0.07 \pm 0.21$ and $\Delta V= 0.06 \pm 0.31$, 
respectively. A systematic offset of 
approximately $-0.13$ mag was observed in the $B-V$ color 
compared to the LEGUS catalog. This discrepancy likely 
stems from the use of average aperture corrections in the 
LEGUS pipeline, which may not adequately account for the 
complex and rapidly varying stellar and nebular backgrounds 
characteristic of active star-forming regions in NGC 1313. 
Notably, \citet{2022MNRAS.510...32D} reported similar 
systematic offsets in LEGUS photometry for NGC 1433, 
suggesting that such discrepancies are often dependent 
on the specific galactic environment and the background 
subtraction methodology employed.

\subsection{Stellar population model}
Simple stellar population models are essential for 
interpreting the integrated light of unresolved stellar 
populations. In this work, \texttt{STARBURST99} 
\citep{1999ApJS..123....3L, 2014ApJS..212...14L} was 
used to derive the underlying IMFs and physical 
parameters of the observed clusters. This model supports 
the parameterization of underlying IMFs. Furthermore, 
the \texttt{Geneva} and \texttt{Padova} stellar 
evolutionary models with different 
metallicities are implemented in the simulations. 
Note that the \texttt{Geneva} models incorporating 
stellar rotation only support two metallicities 
($Z = 0.002$ and 0.014).

We adopted the \texttt{Geneva} model with a 
rotation rate $v_{ini}/v_{break}$ of 0.4, 
which allows us to better interpret the integrated 
spectra of massive stars \citep{2025AJ....169....7K}. 
The synthetic spectra of model clusters were 
generated with age steps of 0.5 Myr for very 
young clusters ($< 15$ Myr), 5 Myr for young 
clusters ($< 100$ Myr), and 50 Myr for older 
clusters ($100$ Myr -- $1$ Gyr). We considered 
the power-law index $\Gamma$ of the 
underlying IMFs, ranging from $-2.5$ to $0.0$ in 
increments of 0.1, for stars with masses greater 
than $0.8 M_{\sun}$. 

NGC 1313 likely hosts subsolar metallicity ($Z = 0.008$) 
environments \citep{2021ApJ...909..121M}. However, the 
synthetic spectra for this metallicity is not supported by 
\texttt{STARBURST99}. The averaged spectra from 
the two sets of the models for $Z = 0.002$ and $0.014$ 
were used as the synthetic spectra for $Z = 0.008$. 
Figure~\ref{fig4} compares synthetic spectra for 
three different metallicities. The integrated spectra 
of extremely young clusters do not show significant 
spectroscopic differences with respect to their 
abundances. A prominent feature is the strength of 
the so-called blue bump, which originates from the 
winds from Wolf-Rayet stars. This spectral feature 
is detectable from 3 Myr to 8 Myr for the solar 
metallicity, but its duration is confined to a much 
shorter timescale at lower metallicities. 

We investigated the relationships between $\Gamma$ 
and some spectral features for $Z = 0.008$ 
(Figure~\ref{fig5}). The slope of the continuum is essentially governed by the content of massive stars. In this study, we measured the slopes of individual synthetic spectra within a wavelength range of 4100 to 5700~\AA\ on a logarithmic scale. Additionally, the Balmer jump appears weak in the spectra of very massive O-type stars, whereas its strength increases significantly for B- and A-type stars, thereby serving as another indicator of the stellar content. We quantified the strength of the Balmer jump as the ratio of the pseudo-continuum flux to the actual fluxes of a given synthetic spectrum between 3650 and 3685~\AA. Lastly, since Wolf-Rayet stars predominantly emerge within an age range of 3 to 6 Myr, the equivalent widths of both the blue and red bumps were measured to provide further chronological constraints.

All relationships essentially 
depend on the cluster ages. At a given age, the 
spectral features vary, to a greater or lesser extent, 
with respect to $\Gamma$. The slope of the continuum 
appears to weakly correlate with $\Gamma$. The strength of the Balmer jump shows 
better sensitivity to the underlying IMF for $\Gamma < -1.5$; 
however, it becomes insensitive for $\Gamma \geq -1.5$. 
The strengths of the blue and red bumps originating 
from Wolf-Rayet stars are likely the best indicator 
of the underlying IMF between 4 Myr and 6 Myr. The 
significance of these relationships are weaker than 
those for the solar metallicity \citep{2025AJ....169....7K}.

It is necessary to constrain the ages of the observed 
clusters; this is essential for determining the underlying IMFs. 
These theoretical relationships serve as a preliminary guide to 
understanding the sensitivity of spectral features to the IMF 
and the potential degeneracies between age and $\Gamma$. While 
they provide physical context for the spectroscopic signatures, 
the final cluster parameters in this study are derived through 
a full-spectrum matching to utilize the complete 
information available in the observed data.

\begin{figure*}[ht!]
\epsscale{1.0}
\plotone{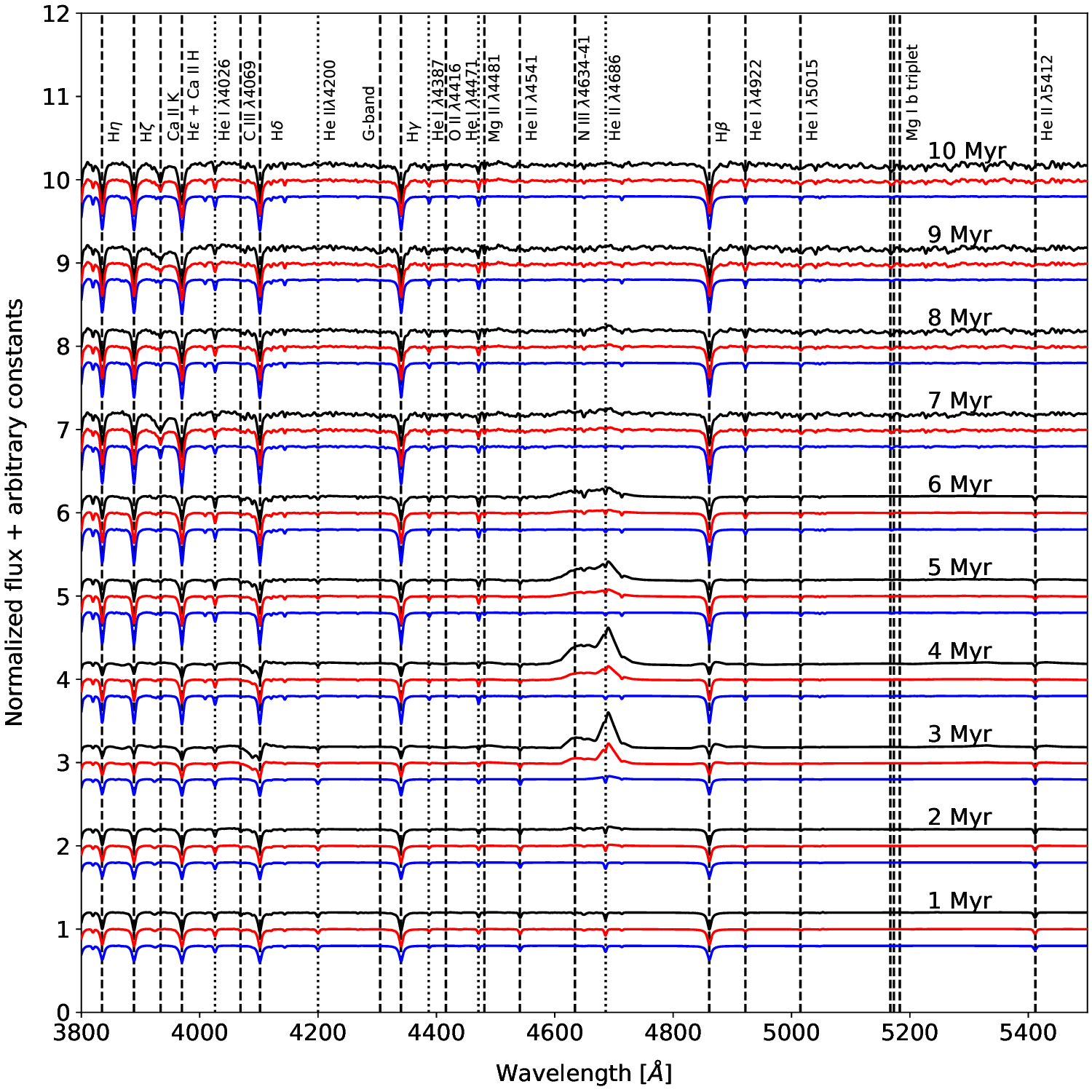}
\caption{Synthetic spectra of the model clusters ($10^6 \ M_{\sun}$) 
for three different metalicities ($Z = 0.002$ -- blue, $0.008$ -- 
red, and $0.014$ -- black). The ages of the clusters are shown 
at the upper-right corner of each spectrum. The Kroupa IMF applied 
to all spectra.}\label{fig4}
\end{figure*}

\begin{figure*}[ht!]
\addtocounter{figure}{-1}
\plotone{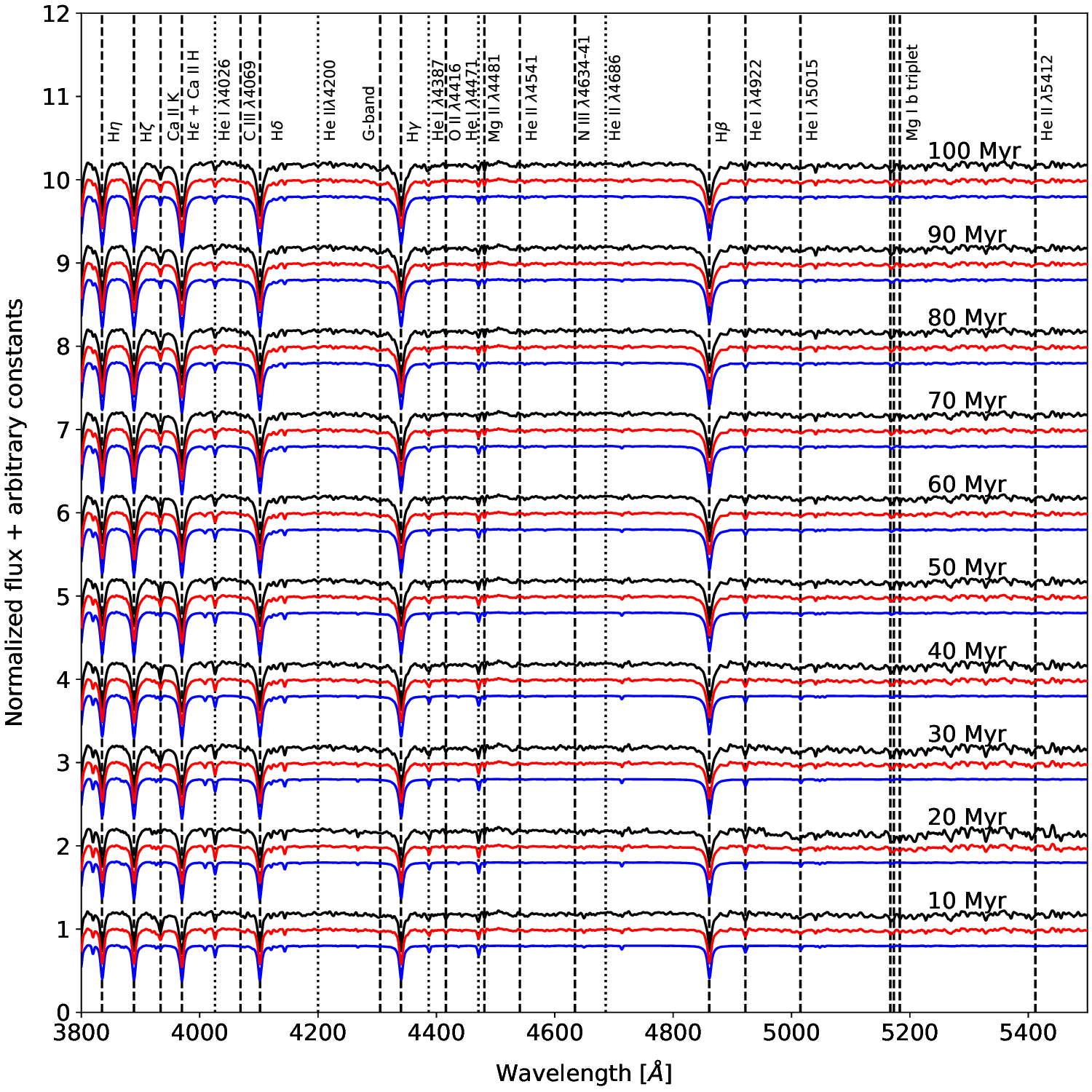}
\caption{Continued.}
\end{figure*}

\begin{figure*}[ht!]
\addtocounter{figure}{-1}
\plotone{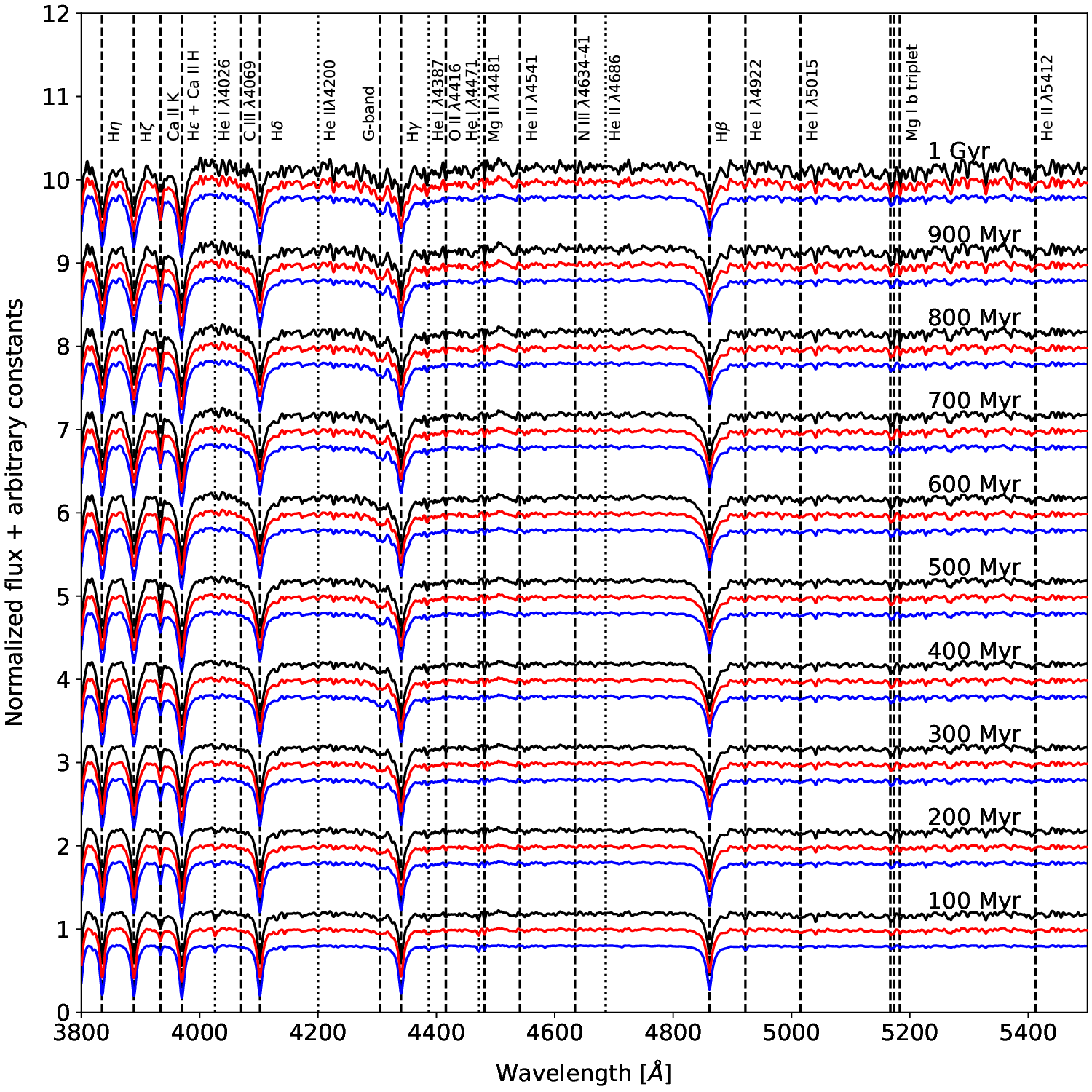}
\caption{Continued.}
\end{figure*}

\begin{figure*}[t]
\epsscale{1.0}
\plotone{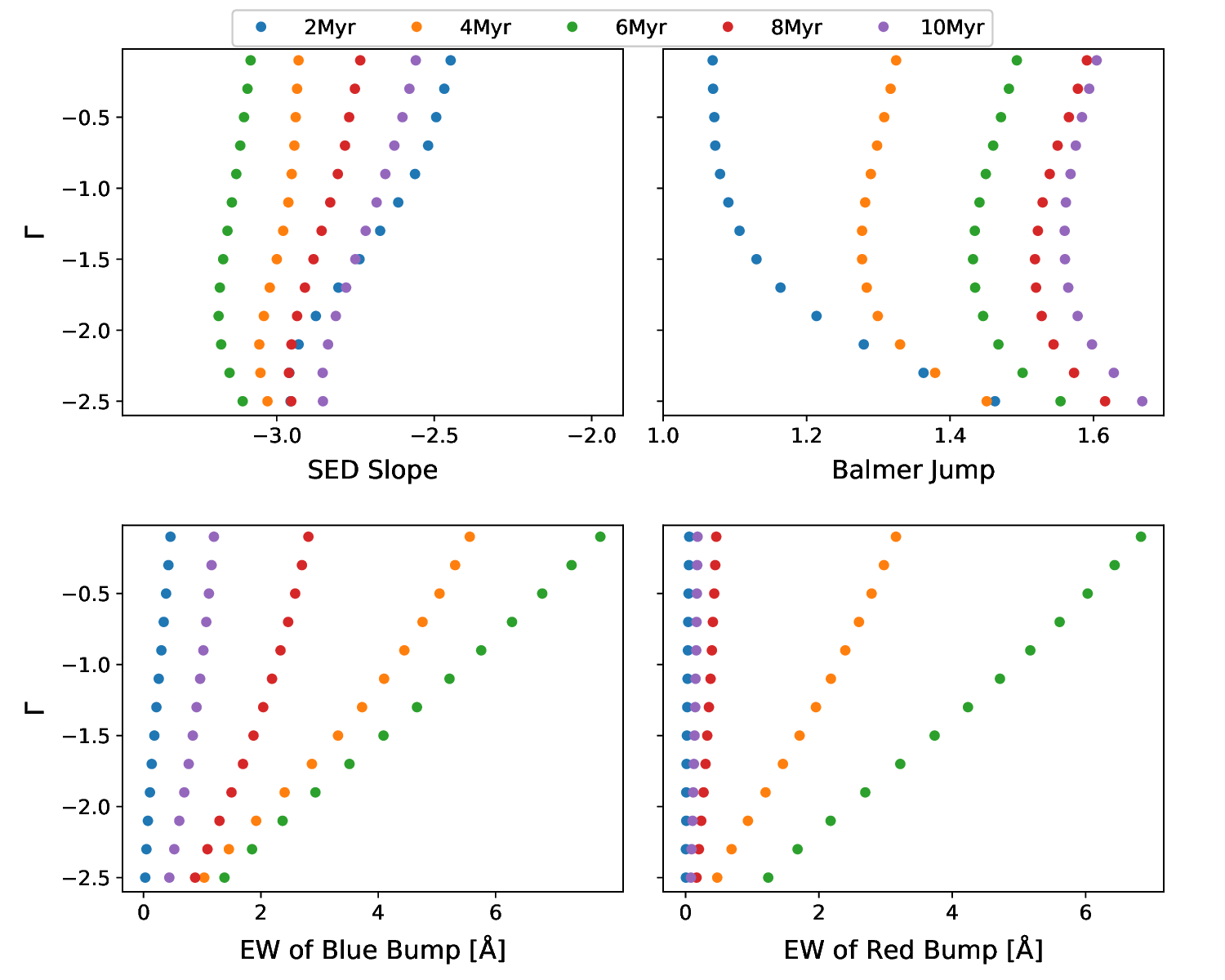}
\caption{Variations of spectral features obtained from 
the synthetic integrated spectra of model clusters ($M_{cl}=10^6 M_{\sun}$ 
and Z = 0.008). In the lower panels, the equivalent widths of blue 
and red bumps are obtained from the output of \texttt{STARBURST99}, 
\ion{N}{3} $\lambda$4640 + \ion{C}{3} $\lambda$4650 
+ \ion{He}{2} $\lambda$4686 and \ion{C}{4} $\lambda$5808, respectively.}\label{fig5}
\end{figure*}

\section{Results} \label{sec:3}
\subsection{Age constraints}\label{sec:31}
The integrated spectral features of clusters are characterized 
by the evolutionary stage of the most massive stars, 
allowing us to infer the ages of the observed 
clusters from their integrated spectra. The synthetic 
spectra are sorted by cluster ages in Figure~\ref{fig4} 
(see the red spectra for $Z = 0.008$). Absorption lines 
are very weak for the first 2 Myr. Contribution of the 
blue bumps to the integrated spectra appears to be 
significant from 3 Myr to 5 Myr. Metallic lines 
strengthen after 7 Myr, while \ion{He}{1} lines diminish. 

The integrated spectra of clusters in an age range from 
10 Myr to 100 Myr do not show a significant variation of 
spectral features with respect to their ages. The strength 
of \ion{Mg}{2} $\lambda$4481 appears to be weaker than 
\ion{He}{1} $\lambda$4471 and it becomes comparable to 
the helium line after 80 Myr. The \ion{Ca}{2} K and 
\ion{He}{1} lines are detectable over the age range. 
A large number of metallic lines strengthen 
in the spectra of older clusters (100 Myr $<$ 
age $\leq$ 1 Gyr), while the \ion{He}{1} lines 
become undetectable. The strength of the G band 
can be used as a useful age indicator. We summarized 
the spectral features related to cluster 
ages as below: 

\begin{enumerate}
\item $\lesssim$ 2 Myr : Very weak spectral lines.
\item 3 -- 5 Myr : The strongest Wolf-Rayet features (blue and red bumps).
\item $<6$ Myr : Detectable \ion{He}{2}$\lambda$5412.
\item $>$ 5 Myr:  Detectable \ion{He}{1} $\lambda\lambda$4387, 4922.
\item $>$ 7 Myr : Detectable \ion{Ca}{2} K, weak \ion{Mg}{2} $\lambda$4481, 
and weak \ion{Mg}{1} b triplet.
\item 10 Myr -- 100 Myr : \ion{Ca}{2} K line and detectable \ion{He}{1} lines.
\item $\gtrsim$ 80 Myr : \ion{Mg}{2} $\lambda$4481 comparable to \ion{He}{1} $\lambda$4471.
\item 100 Myr -- 1 Gyr : G-band, \ion{Ca}{2} K, and \ion{Mg}{1} b triplet 
(systematically age-dependent). Weak \ion{He}{1} lines. \ion{Mg}{2} $\lambda$4481 
stronger than \ion{He}{1} $\lambda$4471.
\end{enumerate}

\begin{figure*}[t]
\epsscale{1.0}
\plotone{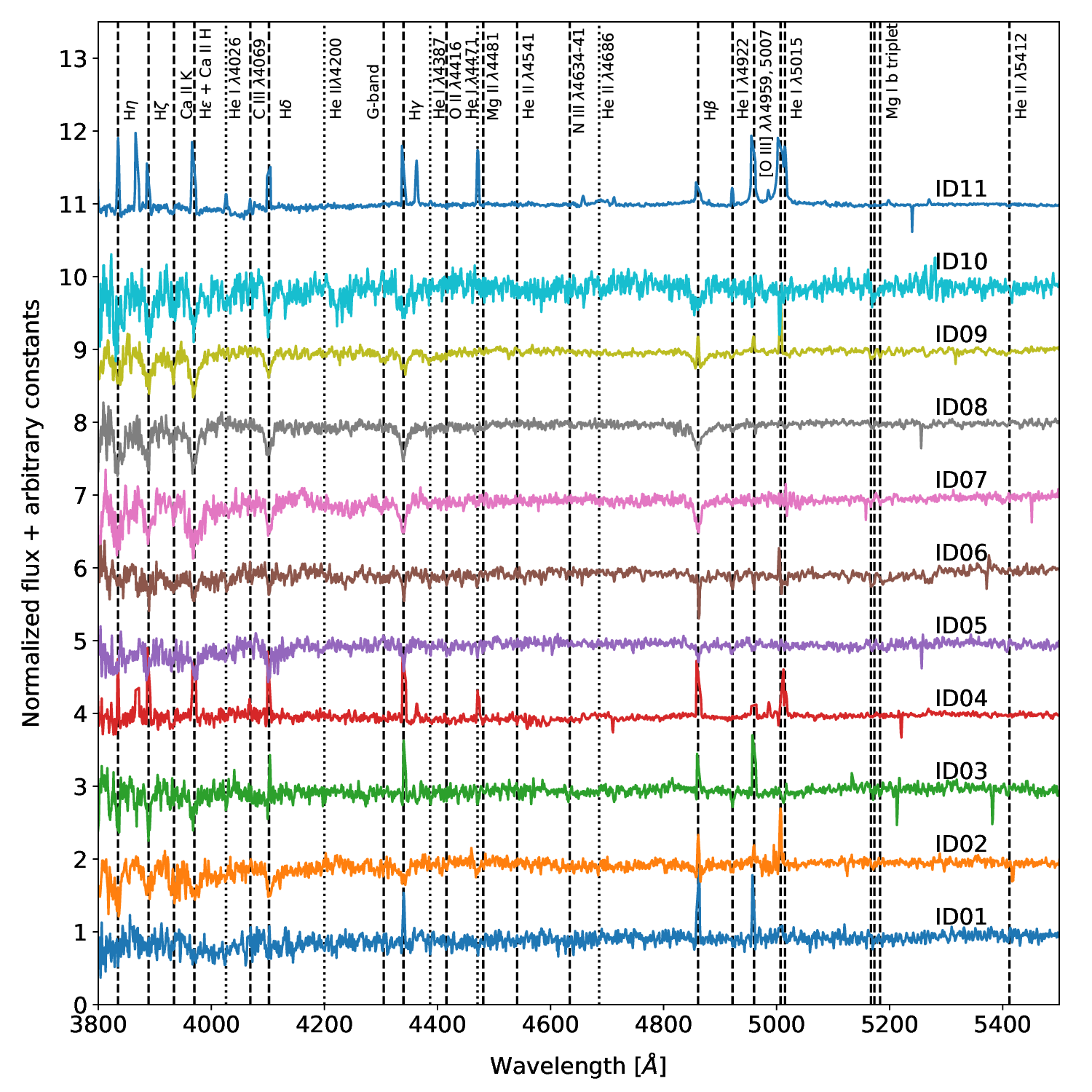}
\caption{Normalized spectra of the observed clusters. 
To minimize spectral overlap, the intensities of strong 
emission lines (exceeding two times the normalized continuum) 
are deliberately limited. The IDs of the individual 
clusters are labeled in the upper-right corner of 
each spectrum. The spectral lines shown in Figure~\ref{fig4} are 
displayed with vertical lines.}\label{fig6}
\end{figure*}

Figure~\ref{fig6} displays the normalized spectra of 
the observed clusters. The clusters ID04 and 11 are 
likely the most youngest clusters among the observed clusters, 
given the strong emission lines as well as the blue 
bump. The existence of the blue bump indicates that 
these clusters are older than 2 Myr and younger than 6 
Myr. The spectra of three clusters (ID01, 02, and 03) also show 
strong emission lines. However, the blue bump is not 
detected in their spectra, which indicates that 
these clusters are either older than 5 Myr or younger than 
3 Myr. \ion{He}{1} lines are clearly seen, whereas 
\ion{He}{2} lines and \ion{Ca}{2} K line are not 
detected. In addition, there is no signal in Mg 
lines. These facts imply that the ages of these 
clusters are about 6--7 Myr. 

The cluster ID05 does not exhibit strong emission 
features, but its absorption line pattern is analogous 
to those of cluster ID01, 02, and 03. Consequently, the age 
of the cluster ID05 is also estimated to be approximately 
6--7 Myr. The spectrum of cluster ID 10 includes 
the \ion{Ca}{2} K absorption line; however, the H$\alpha$ 
and [\ion{S}{2}] $\lambda\lambda$6717, 6731 emission lines 
are also detected (as shown in Figure~\ref{fig2}). These 
emission lines likely originate from winds of massive stars 
\citep{2024ApJ...961...72L} or remaining natal 
clouds \citep{2018MNRAS.477.1993L}. Therefore, the 
age of this cluster may be older than 6 Myr but younger 
than 10 Myr, considering the lifetime of natal clouds 
\citep{1989ApJS...70..731L} and the presence of \ion{Ca}{2} 
K line.

The spectra of the remaining clusters (ID06 -- 09) 
exhibit spectral features characteristic older 
clusters. The \ion{Ca}{2} K line is commonly 
detected in their spectra, reflecting the increasing 
contribution of later-type stars to the integrated 
light. These clusters are, at least, older than 
10 Myr. The detection of the \ion{He}{1} lines 
at 4026, 4387, and 4922 \AA \ suggests younger 
ages than 1 Gyr. The comparison of the \ion{He}{1} 
$\lambda$4471 line with the \ion{Mg}{2} $\lambda4481$ 
yields a tighter constraint on the ages of the 
clusters. The \ion{He}{1} $\lambda$4471 line appears 
to be much stronger than the \ion{Mg}{2} $\lambda4481$ 
in the spectra of the cluster ID06 and 07. Therefore, 
these clusters are likely not older than 80 Myr. 
The comparable strengths of these two spectral 
lines in the spectra of cluster ID08 constrain 
its derived ages to the range of 80 Myr to 100 Myr.

The spectrum of the cluster ID09 exhibits the G-band, 
which suggests the presence of an old stellar 
population. This interpretation is further 
supported by the detection of the very strong 
\ion{Ca}{2} K line. The emission lines 
(H$\beta$ and [\ion{O}{3}] $\lambda\lambda$ 4959, 5007) 
may therefore originate from planetary nebulae. 
The \ion{Mg}{1} b triplet is clearly seen in the 
observed spectrum with a high SNR. In addition, the strength of the \ion{Mg}{2} $\lambda$4481 absorption line becomes stronger than that of the \ion{He}{1} $\lambda$4471 line. Given that the strengths of these two lines are highly comparable within the age range of 100 to 200~Myr, this observed inversion suggests that the age of this cluster is at least 300~Myr.

\subsection{Spectral matching}\label{sec:32}
The integrated light from the observed clusters 
are affected by extinction due to the interstellar medium 
in the Galaxy and their host galaxy. We first applied 
the mean galactic extinction adopting 
$\langle E(B-V)\rangle = 0.09$ in the direction of NGC 1313 \citep{1998ApJ...500..525S,2011ApJ...737..103S} to the 
synthetic spectra for model clusters with a mass of 
$10^6 \ M_{\sun}$. A total-to-selective extinction 
ratio of $R_V = 3.1$ was assumed in the calculation of the 
total extinction $A_V$. 

\begin{figure*}
    \centering
    \includegraphics[width=130mm]{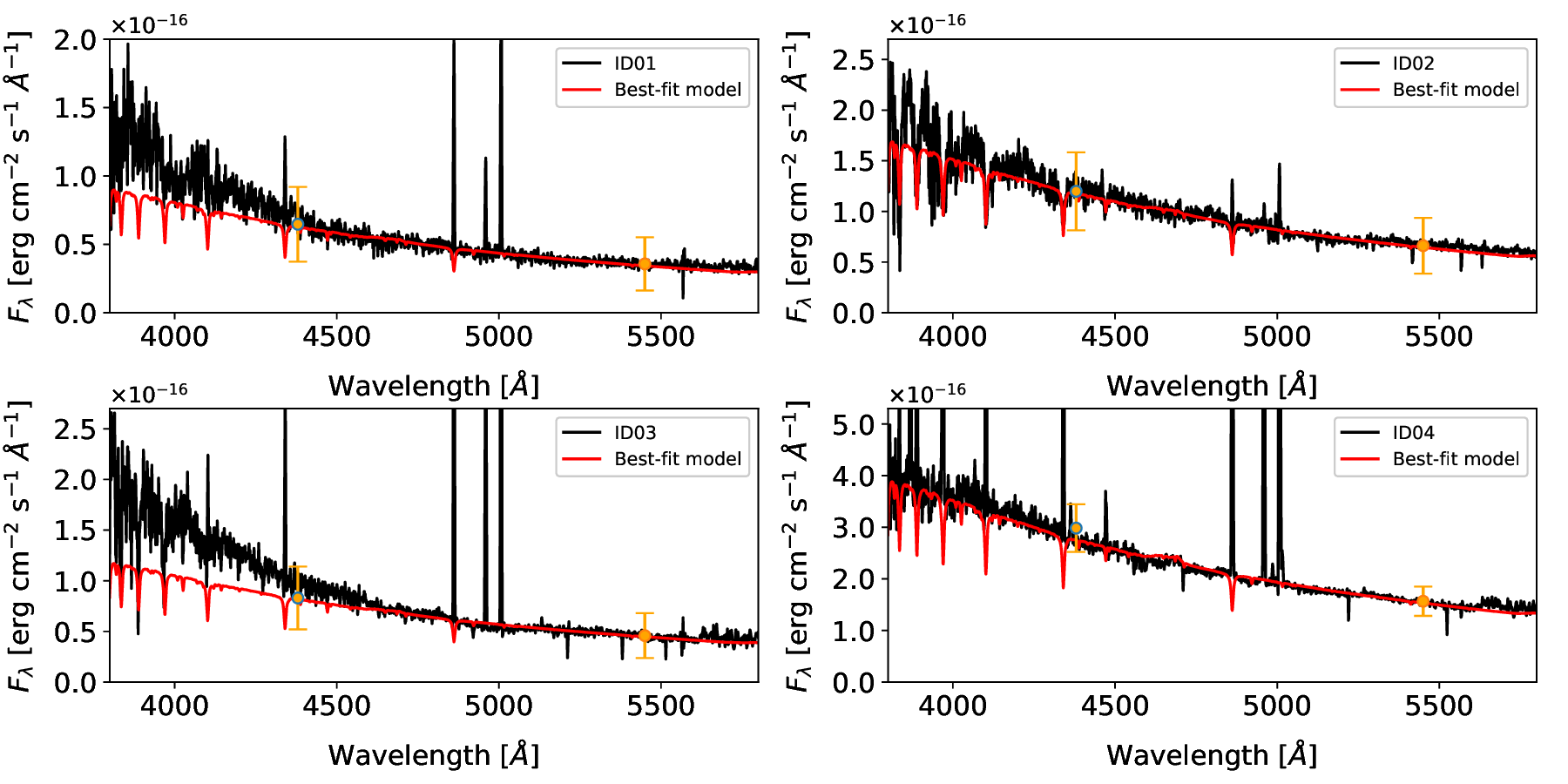}\\
    \includegraphics[width=130mm]{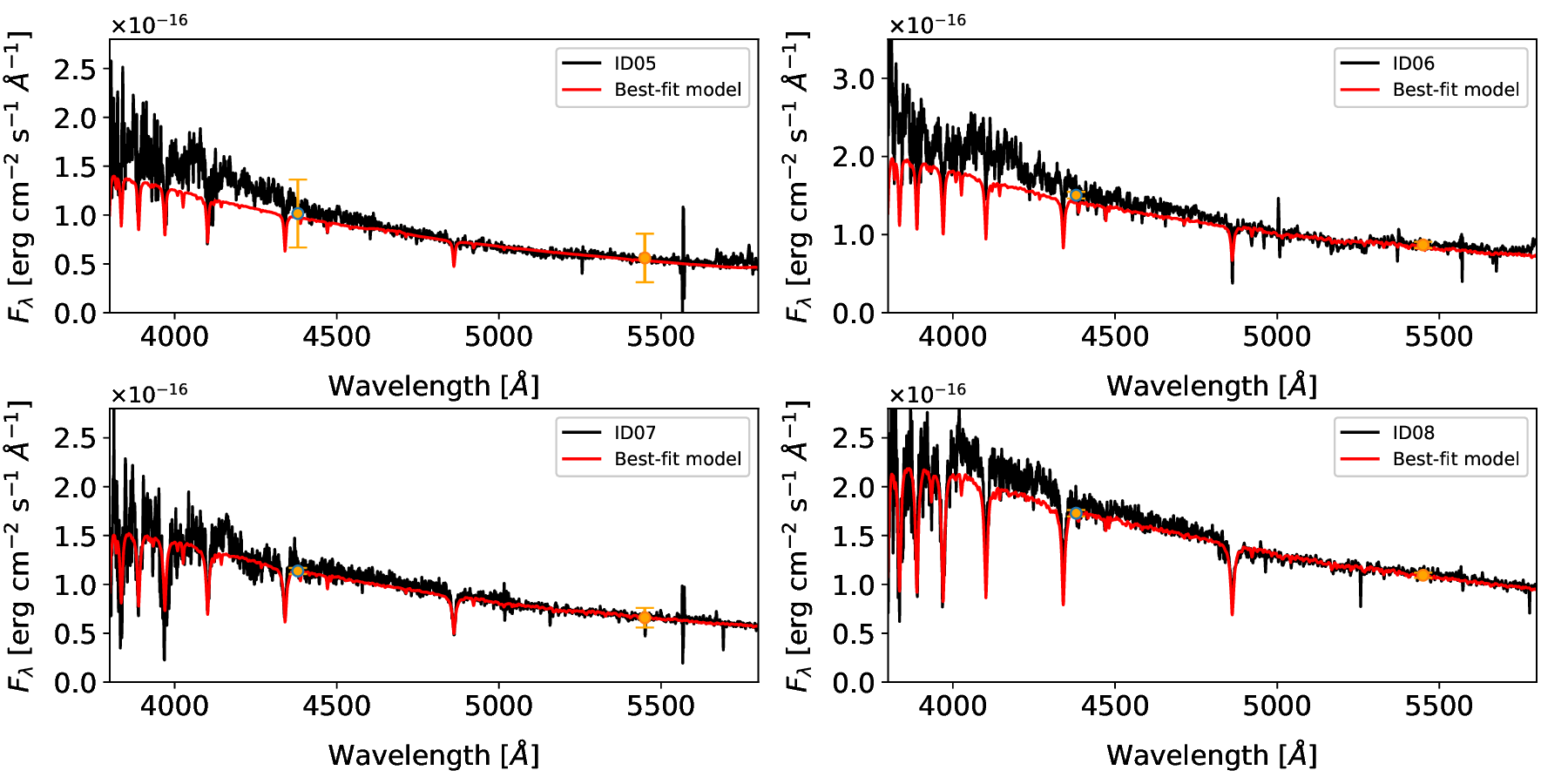}\\
    \includegraphics[width=130mm]{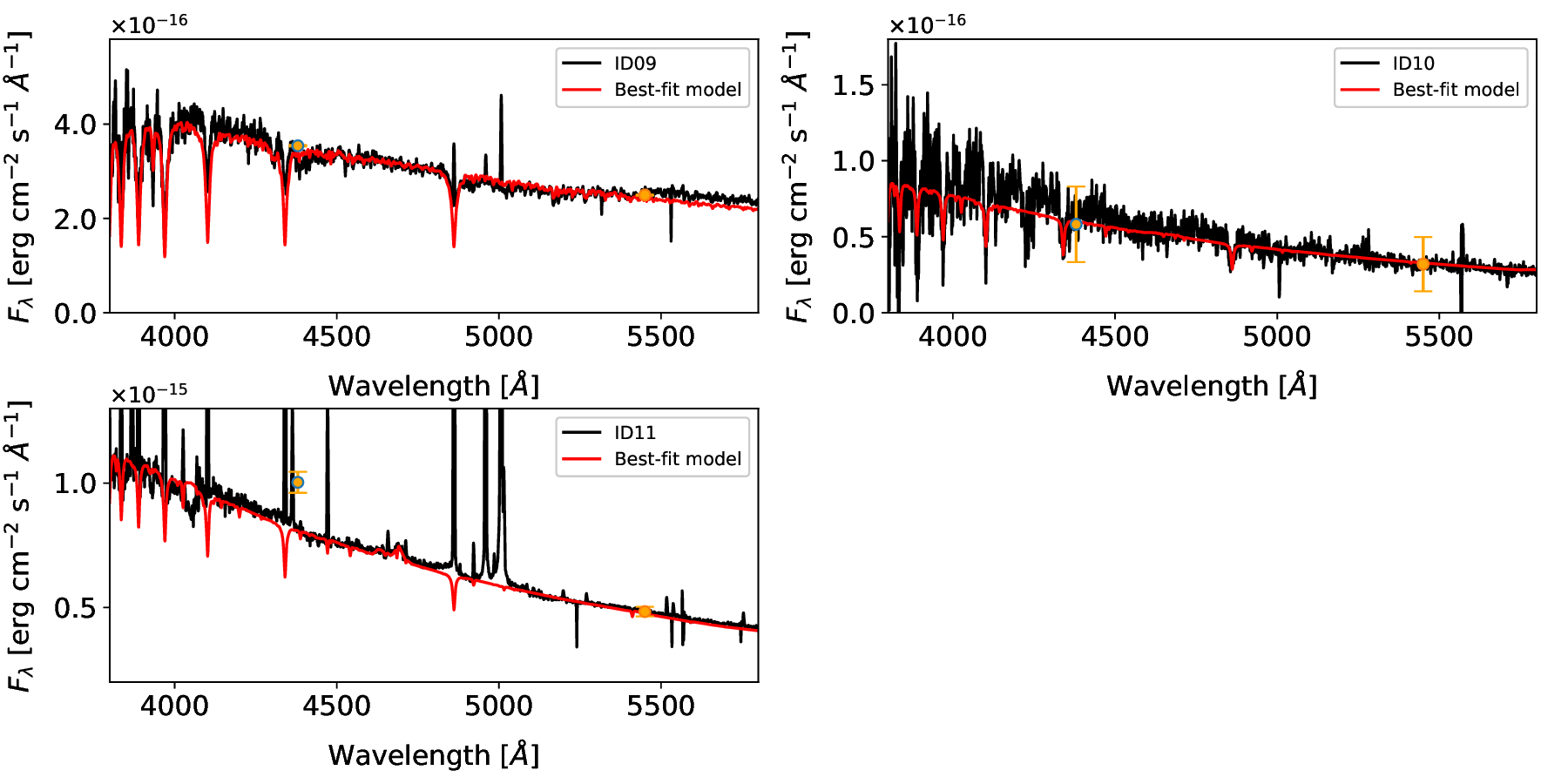}
    \caption{Comparison between the observed spectra (black curves) and the best-matched synthetic spectra (red curves). Note that only the mean Galactic extinction was applied to the synthetic templates. The $B$- and $V$-band mean fluxes predicted by our Monte Carlo simulations are indicated by orange symbols, with the corresponding $1\sigma$ uncertainties denoted by error bars.}
    \label{fig7}
\end{figure*}

\begin{figure}[t]
\epsscale{1.0}
\plotone{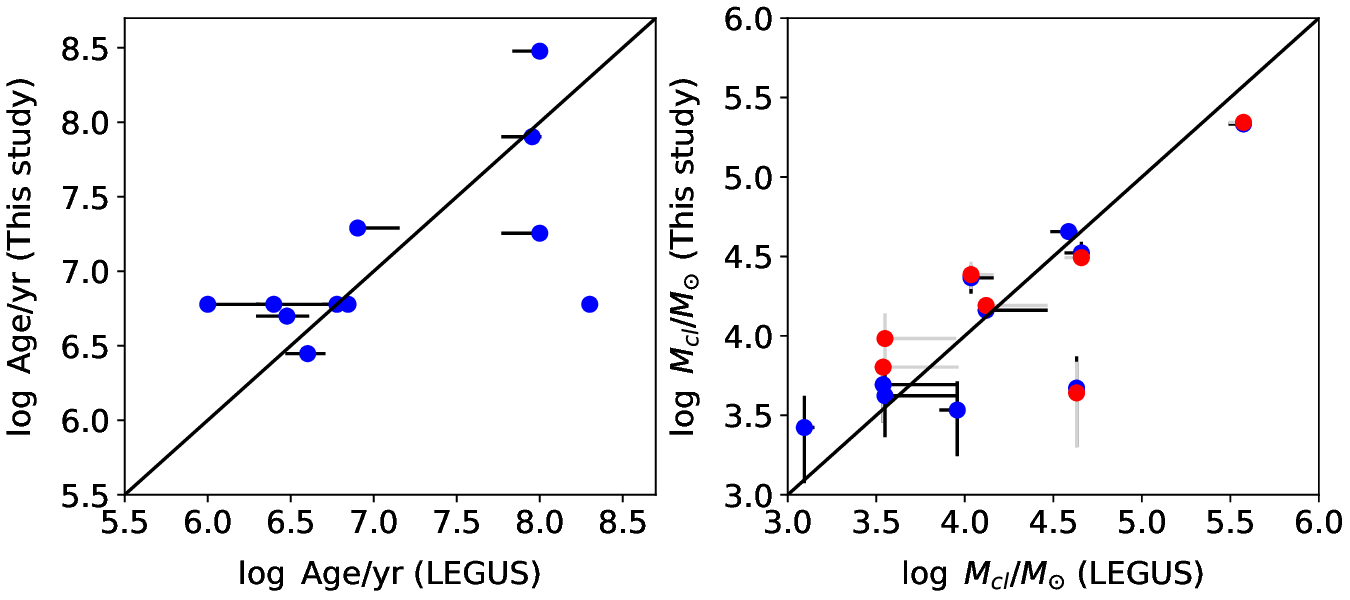}
\caption{Comparison of the physical parameters obtained from this study with those derived in the LEGUS survey \citep{2017ApJ...841..131A}. The left and right panels compare ages and cluster masses estimated in the two studies, respectively. The red symbols represent the results obtained after correcting for internal extinction. In both panels, the horizontal and vertical lines denote the uncertainties associated with each physical parameter.}\label{fig8}
\end{figure}

\begin{figure*}[t]
\epsscale{1.0}
\plotone{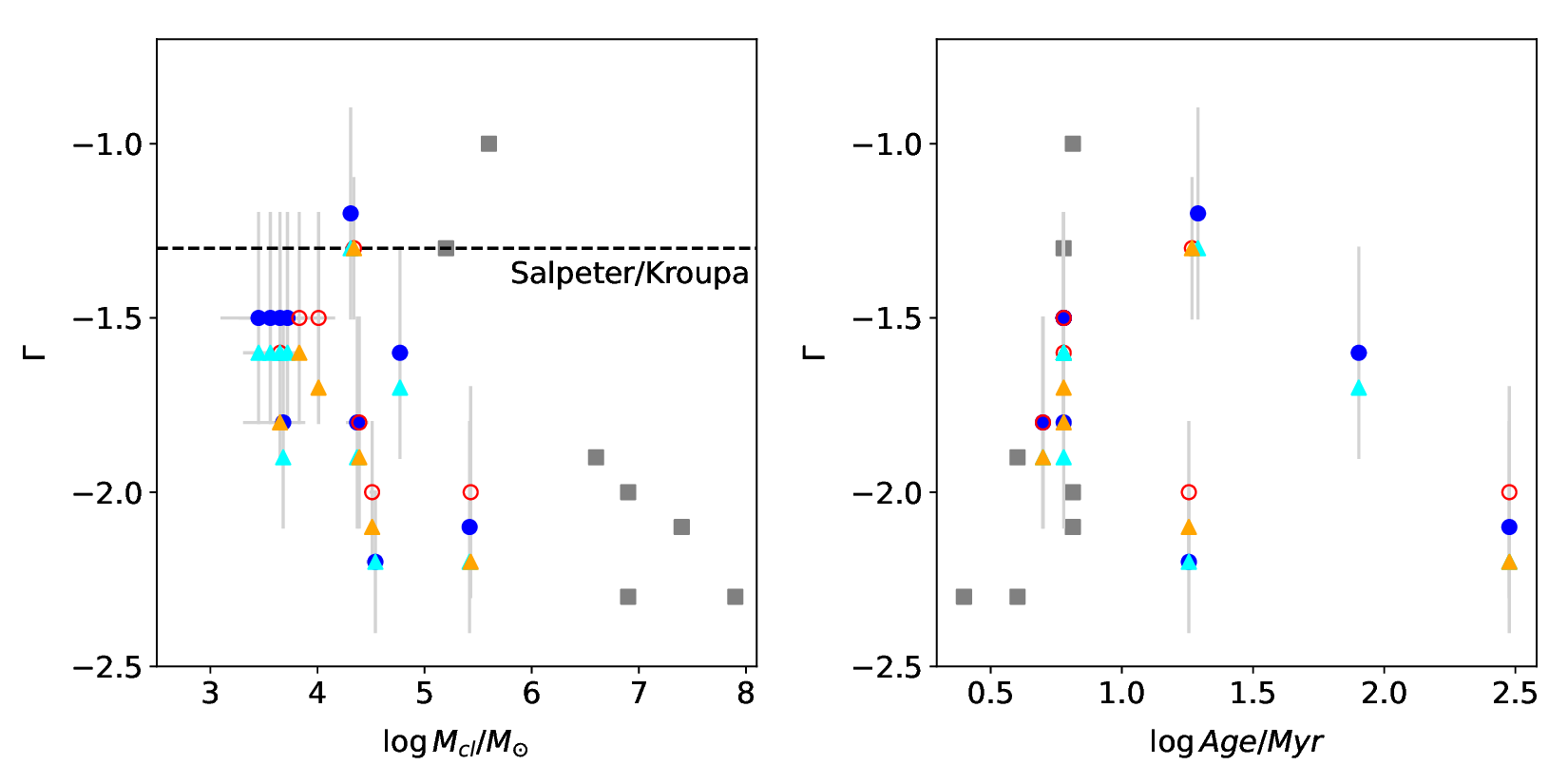}
\caption{Relationship between physical parameters and the IMF slope ($\Gamma$). The left and right panels compare cluster masses and ages against $\Gamma$, respectively. The blue and red circles represent the results obtained without and with internal extinction correction, respectively, while the cyan and orange triangles denote the corresponding best-fit results from the STARBURST99 models. In both panels, the horizontal and vertical solid lines indicate the uncertainties associated with each physical parameter. For comparison, the results for NGC 4038/9 are displayed in gray \citep{2025AJ....169....7K}. The dashed 
line indicates the $\Gamma$ of the Salpeter/Krouap IMFs 
for $\geq 1 M_{\sun}$.}\label{fig9}
\end{figure*}

The synthetic spectra generated from \texttt{STARBURST99} 
were corrected for a distance of 4.4 Mpc \citep{2018ApJS..235...23S} 
and scaled to the continuum levels of individual observed 
clusters between 5000 and 5500 \AA \ (approximately $V$ band). 
This continuum scaling is effective because the flux in $V$ band 
is less subject to the effects of stochastic 
sampling compared to $B$ band. The observed spectra of the 11 
clusters were then compared against a grid of these distance-corrected synthetic spectra. 

The internal extinction estimated from the \ion{Na}{1} 
line (see \citealt{2025AJ....169....7K} for details) and 
age ranges inferred from the observed spectra  were used 
as boundary conditions. We searched for the synthetic counterparts that best reproduce the observed spectra by minimizing the residuals between the data and the models. The residual is defined as the mean squared error over the given wavelength range: 

\noindent $$\text{Residual} = \frac{1}{N} \sum_{i=1}^{N} (O_i - M_i)^2$$

\noindent where $O_i$ and $M_i$ represent the fluxes of the observed spectrum and the synthetic model at each wavelength, respectively, and $N$ is the total number of wavelength points. This procedure was conducted within a spectral window between 3800~\AA\ and 5800~\AA, where prominent emission lines were masked out, as the SNRs of the observed spectra degrade significantly at wavelengths shorter than 3800~\AA. Figure~\ref{fig7} displays the 
observed spectra with the best-fit synthetic spectra.  

With the exception of the cluster ID05, the internal 
reddening $E(B-V)$ derived from spectral matching 
ranges from 0.0 to 0.09, indicating negligible 
extinction. We assumed zero internal extinction 
for all clusters. For most clusters, the slopes 
of their blue spectra appear to be even steeper than 
those of the best-fit unreddened synthetic spectra. 
This discrepancy likely stems from the effects of 
stochastic sampling given that clusters have 
low stellar masses ($< 10^4 \ M_{\sun}$; e.g., 
\citealt{2017ApJ...841..131A}). To test this effect, 
we conducted simulations based on the Monte Carlo 
method. The cluster ages and $\Gamma$ values obtained 
from the best-fit synthetic spectra of the simple 
stellar population model were adopted in our 
simulations. The masses of individual stars 
were generated from the IMFs characterized by 
the $\Gamma$ values. We considered 
four model clusters with total initial masses of 
about $10^3$, $10^4$, $10^5$, and $10^6 \ M_{\sun}$ 
for given ages and the underlying IMFs.

The survival status of the most massive stars 
is inherently dictated by the presence of the most 
massive remaining star on the Geneva isochrones 
\citep{2012A&A...537A.146E,2013A&A...558A.103G} 
at the given ages. The absolute magnitudes of 
remaining stars in $B$ and $V$ bands were obtained 
by interpolating their initial masses to the 
mass-luminosity relations of the adopted 
isochrones for $Z = 0.014$ and $0.002$, 
respectively. We computed their apparent magnitudes 
at the distance of 4.4 Mpc after correction for the 
Galactic mean extinction. The bandpass-averaged fluxes 
in $B$ and $V$ bands were obtained using \texttt{get\_flux} which 
is a \texttt{pyphot} module \citep{fouesneau2025pyphot}, respectively. 
Finally, we computed the total fluxes by summing the fluxes of individual stars, and then averaged these values obtained from the Geneva models for $Z = 0.014$ and $0.002$. This procedure was repeated 10,000 times for given total cluster masses.

The mean fluxes and their standard deviation 
values over 10,000 trials for model clusters with 
different initial masses ($\sim 10^3 - 10^6 \ M_{\sun}$) 
were calculated. The simulated fluxes in $B$ and 
$V$ were then scaled to the observed spectra by anchoring 
them to the observed fluxes in the $V$ band. Figure~\ref{fig7} compares the 
mean fluxes inferred from our simulations (orange symbols) 
with the observed spectra. 

STARBURST99 generates integrated spectra by scaling 
a continuous IMF based on the total cluster luminosity. 
Consequently, in the low-cluster-mass regime where 
the total number of member stars is limited, the model 
effectively includes fractional contributions from 
massive stars (i.e., less than one star), which can 
lead to an underestimation of the model flux compared 
to the observed $B$-band spectra. Notably, while our 
simulations align with the STARBURST99 predictions 
on average, the observed spectrum is well-explained 
with the 1-$\sigma$ uncertainty of the simulated flux 
distribution, with the exception of ID11. 

The simulated $B$-band flux of ID11 appears to be greater than the observed flux. Since this cluster was not spatially resolved under our seeing-limited conditions, its spectrum suffered from contamination by neighboring sources. Consequently, the total luminosity of the cluster was overestimated, and the resulting continuum slope became atypical of a coeval stellar population. Because the best-fit STARBURST99 model was selected based on this artificially elevated $V$-band normalization, the grid search was biased toward an unphysical template. As a result, subsequent Monte Carlo simulations based on these skewed parameters yield an enhanced $B$-band flux that systematically exceeds both the empirical data and the STARBURST99 prediction. We could not derive any other physical parameters for this cluster, only providing a constraint on its age.

We identified the top ten unreddened synthetic spectra that best matched the observed spectra for each cluster. The physical parameters of these ten models were then averaged, and their standard deviation values were adopted as uncertainties for given parameters. The cluster masses and the numbers of member stars were determined by interpolating the observed $V$-band fluxes into the respective relationships derived from our simulations—specifically, the relation of $V$-band fluxes to cluster masses and to total star counts. The associated uncertainties for both parameters were then defined by converting the $1\sigma$-fluctuations of the simulated $V$-band fluxes into their corresponding physical units. The same procedure was repeated for reddened synthetic spectra, allowing us to examine the effect of internal extinction correction on the results. Our results are summarized in Table~\ref{tab1}.

\begin{deluxetable*}{lccccccccccc}
\tiny
\rotate
\tablewidth{0pt}
\setlength{\tabcolsep}{2pt}
\tablecaption{Physical properties of 11 clusters in NGC 1313 \label{tab1}}
\tablehead{
\colhead{ID} & \colhead{R.A.} & \colhead{decl.} &  
\colhead{Age} & \colhead{$\log M_{\mathrm{cl}}/M_{\sun}$} & \colhead{$N_{\star}$}& \colhead{$\Gamma$} &
\colhead{$E(B-V)$} & \colhead{Age} &  \colhead{$\log M_{\mathrm{cl}}/M_{\sun}$} & \colhead{$N_{\star}$} &\colhead{$\Gamma$} \\
 & \colhead{[Deg]} & \colhead{[Deg]} &  
\colhead{[Myr]} &  &&  &
& \colhead{[Myr]} &&   & }
\startdata
 1 & 49.538176 & -66.50585 & $6.0$ & $3.42^{+0.19}_{-0.34} \ (3.45^{+0.19}_{-0.34})$ & $1405\pm767$& $-1.5\pm0.3 \ (-1.6)$ & 0.0 &  \nodata & \nodata & \nodata & \nodata \\
 2 & 49.554503 & -66.499382 & $6.0$ & $3.69^{+0.15}_{-0.23} \ (3.72^{+0.15}_{-0.23})$ & $2606\pm1084$& $-1.5\pm0.3 \ (-1.6)$ & 0.09 & $6.0$ & $3.80^{+0.04}_{-0.34} \ (3.83^{+0.15}_{-0.20})$ & $3368\pm1287$& $-1.5\pm0.3 \ (-1.6)$\\
 3 & 49.585968 & -66.478991 & $6.0$ & $3.53^{+0.17}_{-0.28} \ (3.56^{+0.17}_{-0.28})$ & $1803\pm872$& $-1.5\pm0.3 \ (-1.6)$ & 0.00 & \nodata  &  \nodata & \nodata & \nodata\\
 4 & 49.598377 & -66.479833 & $5.0$ & $4.37^{+0.07}_{-0.09} \ (4.37^{+0.07}_{-0.09})$ & $13984\pm2549$& $-1.8\pm0.3 \ (-1.9)$ & 0.02 & $5.0$ & $4.39^{+0.07}_{-0.08} \ (4.39^{+0.07}_{-0.08})$  & $14835\pm2591$& $-1.8\pm0.3 \ (-1.9)$\\
 5 & 49.642390 & -66.483353 & $6.0$ & $3.62^{+0.16}_{-0.26} \ (3.65^{+0.16}_{-0.25})$ &$2207\pm979$& $-1.5\pm0.3 \ (-1.6)$ & 0.22 & $6.0$ & $3.98^{+0.15}_{-0.20} \ (4.01^{+0.14}_{-0.21})$  & $5382\pm2039$& $-1.5\pm0.3 \ (-1.7)$\\
 6 & 49.649185 & -66.489564 & $19.5\pm1.5$ & $4.16^{+0.02}_{-0.02} \ (4.31^{+0.02}_{-0.02})$ & $7515\pm342$& $-1.2\pm0.3 \ (-1.3)$ & 0.03 & $18.5\pm2.3$ & $4.19^{+0.02}_{-0.02} \ (4.34^{+0.02}_{-0.02})$& $8175\pm354$& $-1.3\pm0.2 \ (-1.3)$\\
 7 & 49.627350 & -66.503816 & $18\pm2.4$ & $4.52^{+0.06}_{-0.07} \ (4.54^{+0.06}_{-0.07})$ & $23635\pm3591$ & $-2.2\pm0.2 \ (-2.2)$ & 0.02 & $18\pm2.4$ & $4.49^{+0.02}_{-0.01} \ (4.51^{+0.02}_{-0.01})$& $21500\pm902$ & $-2.0\pm0.2 \ (-2.1)$\\
 8 & 49.608648 & -66.493019 & $80$ & $4.66^{+0.01}_{-0.01} \ (4.77^{+0.01}_{-0.01})$ & $31202\pm752$& $-1.6\pm0.3 \ (-1.7)$ & 0.00 &\nodata  & \nodata & \nodata & \nodata\\
 9 & 49.569284 & -66.494668 & $300$ & $5.33 \ (5.42)$ &$179869\pm773$& $-2.1\pm0.3 \ (-2.2)$ & 0.01 & $300$ & $5.34 \ (5.43)$ & $185081\pm790$& $-2.0\pm0.3 \ (-2.2)$\\
10 & 49.517642 & -66.513214 &  $6.0$ & $3.67^{+0.19}_{-0.36} \ (3.68^{+0.19}_{-0.36})$ &$2847\pm1583$& $-1.8\pm0.3 \ (-1.9)$ & 0.06 & $6.0$ & $3.64^{+0.19}_{-0.33} \ (3.65^{+0.19}_{-0.33})$ &$2546\pm1354$& $-1.6\pm0.3 \ (-1.8)$\\
11 & 49.512998 & -66.503938 & $2.8\pm0.2$ & \nodata & \nodata & \nodata & \nodata & \nodata & \nodata & \nodata & \nodata \\
\enddata
\tablecomments{Column (1) : The IDs of individual clusters. Columns (2) and (3) : The equatorial coordinates (J2000). Columns (4) -- (7) : The age, cluster mass, the number of stars, and $\Gamma$ derived from this study assuming the zero extinction. Values in parentheses indicate the initial cluster masses and the best-fit $\Gamma$ results, respectively. Column (8): Reddening estimated from the \ion{Na}{1} absorption feature. Columns (9)--(12): Same as Columns (4)--(7), but derived after correcting for internal extinction.}
\end{deluxetable*}

\subsection{Physical parameters}\label{sec:33}
The ages of the observed clusters range from 
2 Myr to 6 Myr for extremely young clusters, 
and the age of the oldest cluster is 
approximately 300 Myr. The other clusters 
are younger than 100 Myr. The most massive 
cluster among our targets contains a total 
stellar mass of $2.6 \times 10^5 M_{\sun}$, 
whereas the stellar masses of the lightest 
cluster is $2.8 \times 10^3 M_{\sun}$. The 
power-law index $\Gamma$ of the IMF was found 
in a range from $-2.2$ to $-1.2$.

We confirmed internal consistency among 
the physical parameters derived from the top 
ten models, which all yielded consistent ages 
for a given cluster. The standard deviation of 
the power-law index $\Gamma$ across these top 
ten models is approximately 0.3. When restricting 
the analysis to the top three models, this 
typical error narrows to about 0.1. Furthermore, 
the underlying IMFs, ages, and cluster masses 
remain consistent within their associated errors, 
regardless of the extinction correction applied.

Figure~\ref{fig8} presents a comparison between 
the ages and stellar masses derived in this study 
and those reported by the 
LEGUS survey \citep{2017ApJ...841..131A}. The logarithmic 
ages obtained here are, on average, consistent with the LEGUS 
values, with a mean difference of $\Delta \log (\mathrm{Age/yr}) 
= -0.03 \pm 0.61$ (where $\Delta$ denotes the offset 
between this study and LEGUS). The cluster masses derived 
in this study are, on average, in good agreement with the 
LEGUS estimates. The mean difference between the two studies 
is less than 0.1 dex, regardless of the applied extinction correction.

The underlying IMFs of most clusters appear to 
deviate from the standard Salpeter/Kroupa IMF ($\Gamma = -1.3$) 
\citep{1955ApJ...121..161S,2001MNRAS.322..231K} and 
exhibit a correlation with cluster mass (the left panel 
of Figure~\ref{fig9}). 
Specifically, more massive clusters tend to be 
characterized by a smaller $\Gamma$. This trend 
remain consistent when using result derived from 
extinction-corrected synthetic spectra (red dots). 

A similar relationship has been reported for several 
young clusters in the Antennae Galaxies (NGC 4038/9) 
\citep{2025AJ....169....7K}. In this study, we compared 
our findings with the relationship established 
in the aforementioned work (indicated by gray squares). 
The clusters in NGC 4038/9 are generally more massive than 
those in NGC 1313. Despite this difference in mass scale, 
the relationships in both galaxy systems are 
remarkably similar. 

To assess whether the observed trends are influenced 
by the intrinsic age-IMF degeneracy, we examined the relationship 
between the derived cluster ages and $\Gamma$ values, as presented 
in the right panel of Figure~\ref{fig9}. We found no systematic 
correlation between age and $\Gamma$ for the clusters in either 
NGC 1313 or NGC 4038/9. The lack of a clear trend indicates 
that the IMF slope variations are statistically independent 
of the cluster age in our sample. Therefore, it is unlikely 
that the reported mass-IMF relationship is a byproduct of 
the age-IMF degeneracy or observational selection effects related 
to age.

\begin{figure*}
    \centering
    \includegraphics[width=130mm]{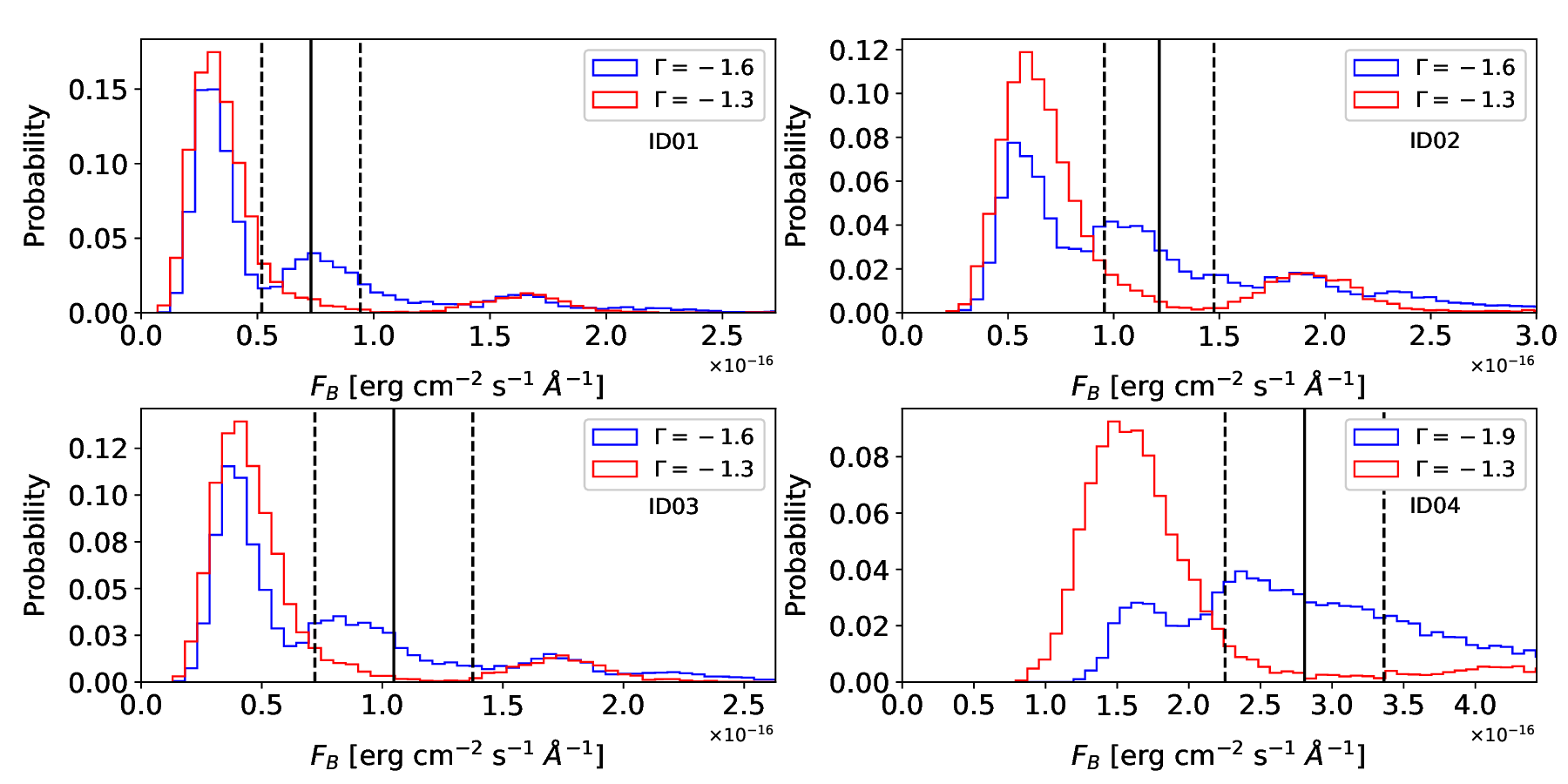}\\
    \includegraphics[width=130mm]{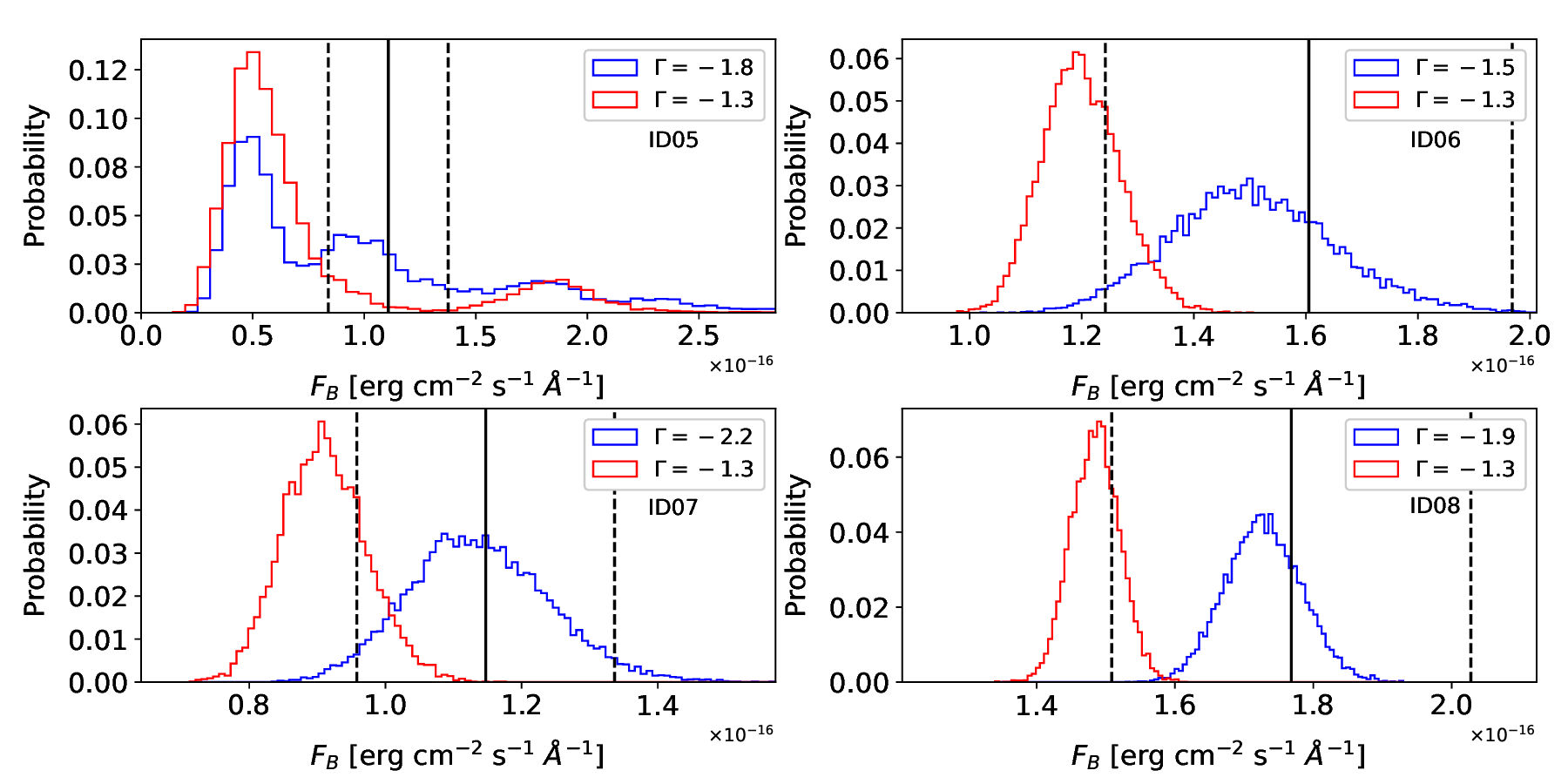}\\
    \includegraphics[width=130mm]{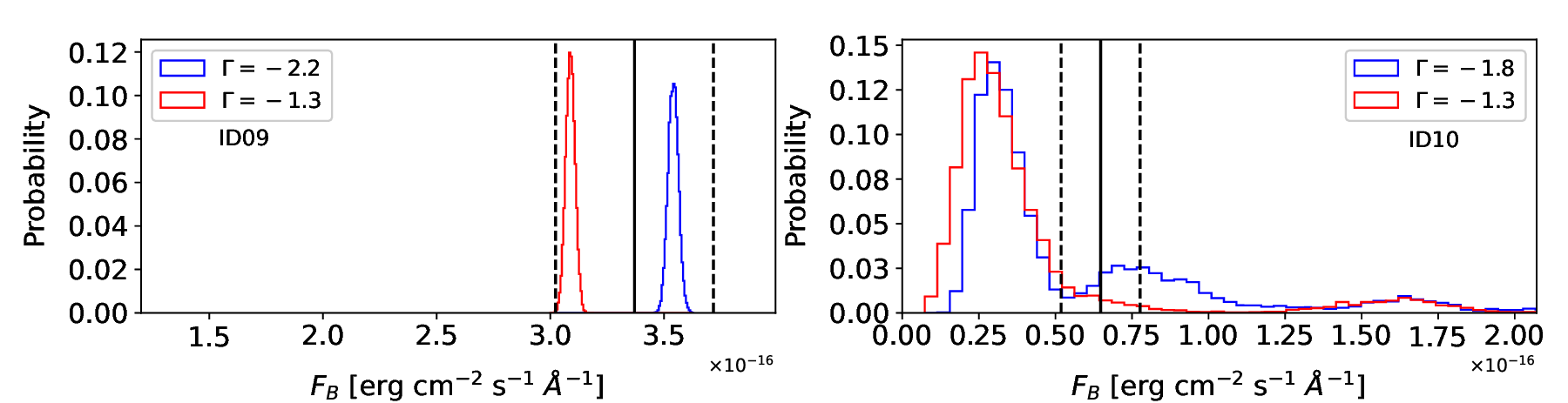}
    \caption{Flux distributions in the $B$-band generated from 10,000 Monte Carlo simulations for the ten observed clusters. Each panel represents an individual cluster, displaying the simulated histograms obtained using the IMF slope ($\Gamma$) from this study (blue) and under the assumption of a standard Salpeter/Kroupa IMF (red). The vertical solid line indicates the observed mean $B$-band flux of the spectrum, while the vertical dashed lines denote its $1\sigma$ uncertainties.}
    \label{fig10}
\end{figure*}

\section{Discussion} \label{sec:4}
We derived the physical parameters of ten stellar clusters that likely formed under subsolar metallicity in the barred galaxy NGC 1313. The observed clusters represent a bright population within their host galaxy, with stellar masses ranging from $2.8 \times 10^3 M_{\sun}$ to $2.6 \times 10^5 M_{\sun}$. The underlying IMFs of these clusters tend to deviate from the standard Salpeter/Kroupa IMF, exhibiting a noticeable correlation between the power-law index $\Gamma$ and their masses. Here, we discuss potential factors that could affect our results.

The first factor is the effect of the spectral SNRs on the derived physical parameters of the clusters. The SNRs of the observed spectra, measured at 5400~\AA, range from 16 to 137. Although Cluster ID10 exhibits the lowest SNR, its physical parameters derived from the top three models are distributed within narrow ranges ($\pm 0.1$), which is highly consistent with the behaviors observed in the spectra of other clusters with higher SNRs. This fact suggests that the SNRs of our empirical spectra are unlikely to induce systematic variations in the underlying IMFs with respect to cluster masses.

The second factor is the effect of stochastic 
sampling. Massive stars are rare in low-mass 
clusters, and therefore the existence of massive 
stars exerts a profound influence on the resultant 
integrated spectrum, often dominating its overall 
characteristics \citep{2013NewAR..57..123C}.
The integrated spectra and photometry of 
clusters containing total stellar masses smaller 
than $10^4 M_{\sun}$ are subject to this 
statistical effect \citep{2015MNRAS.452.1447K}.
In addition, the IMFs derived from a star 
count method exhibit a large dispersion for 
clusters with a small number of members ($< 3000$) \citep{2015AJ....149..127L}. 

A total of five clusters (ID01, 02, 03, 05, and 10) in our targets 
contain stellar masses smaller than 
$10^4 M_{\sun}$, and therefore they are possibly 
affected by the effect of stochastic sampling. 
Taking this effect into account, the observed 
spectra were simply compared with the mean fluxes 
derived from the Monte Carlo simulations in Section~\ref{sec:32}. We further 
investigated the effect of stochastic sampling through 
the distributions of the $B$-band fluxes in detail. 
Based on the best-fit parameters and the number of 
members summarized in Table~\ref{tab1}, a total of 
10,000 model clusters were generated from the Monte Carlo 
simulations for a given observed cluster. The $B$-band 
fluxes were calculated by using the same method described 
in Section~\ref{sec:32}.

Figure~\ref{fig10} displays the distributions of the simulated $B$-band fluxes (blue histograms) for our target clusters. A prominent peak is observed at a low-flux level for the low-mass clusters, indicating the most probable spectral flux expected during observations. In addition, a secondary peak and an extended tail appear toward higher flux levels, representing the flux generated when massive stars are stochastically included. These results strongly suggest that the low-mass clusters in our sample are highly subject to stochastic sampling effects. On the other hand, the more massive clusters (ID06, 07, and 08) exhibit broad flux distributions but lack the multiple discrete components characteristic of their lower-mass counterparts.

If the blue excess in the observed spectra of the low-mass clusters is the effect of stochastic sampling, one can speculate why clusters having spectra with the most probable fluxes were not observed. It is likely due to an observational selection bias. The least massive clusters in our sample have an apparent magnitude of $V \sim 20$ mag, which represents the observational limit barely accessible even with 8-meter-class telescopes for this study. If these clusters had not stochastically sampled additional massive stars, their intrinsic luminosities would have been significantly fainter, dropping them below our detection threshold. Consequently, such clusters would have been systematically excluded from our target selection, leaving only the stochastically brightened counterparts observable. 

Thirdly, if the mass distribution of a low-mass cluster originates from a standard Kroupa IMF, it will likewise exhibit stochastic sampling effects. In such cases, it is necessary to test whether the true parent population responsible for the observed flux is the standard Salpeter/Kroupa IMF ($\Gamma = -1.3$) or the underlying IMF characterized by our best-fit $\Gamma$. To address this, we performed the same simulations using a fixed $\Gamma$ of $-1.3$. The number of cluster members for this control run was deduced by anchoring the simulated $V$-band fluxes to the observed values. The resulting distributions are displayed as the red histograms in Figure~\ref{fig10}.

For the low-mass clusters (ID01, 02, 03, 05, and 10), while the first strong peak at a low-flux level aligns well between both sets of simulations, the location of the secondary peak shows a noticeable discrepancy. When comparing the observed $B$-band flux and its associated 1-$\sigma$ uncertainty against these distributions, the observed data point falls squarely within the secondary peak of the distribution generated by our derived $\Gamma$. In contrast, the simulation assuming a standard Salpeter/Kroupa IMF ($\Gamma = -1.3$) places this second peak at substantially higher flux levels. Although we cannot entirely exclude the Salpeter/Kroupa IMF as a viable parent population based purely on these numerical statistics, the tight alignment of the observational data suggests that the observed clusters preferentially favor the underlying IMF characterized by our derived $\Gamma$.

For the higher-mass clusters (ID04, 06, 07, and 08), both sets of simulations commonly display broad, single-peaked flux distributions with substantial overlap, contrasting with the bimodal distributions seen in the lower-mass counterparts. This behavior indicates that stochastic sampling effects are partially mitigated as cluster mass increases. However, these stochastic effects are not entirely negligible; in fact, the observed $B$-band fluxes for some clusters are slightly higher than the values predicted by the STARBURST99 models. Overall, the observed $B$-band fluxes are located closer to the distributions derived from our results. For the cluster ID09, the simulated $B$-band fluxes from both cases exhibit narrow distributions, and the observed flux lies between these two primary distributions, though slightly closer to the blue histogram derived from our results. Crucially, both simulated flux distributions are well-contained within the $1\sigma$ uncertainty window of the observed flux, leaving the underlying IMF of this cluster somewhat uncertain. Given its advanced age of approximately 300 Myr, both dynamical and stellar evolution may, in fact, have erased the imprints of the parent mass distribution.

We have comprehensively addressed the potential systematic and probabilistic factors that could artificially induce or influence our results. Foremost, the constraints imposed by the spectral SNR were evaluated to ensure the robustness of our parameter derivation. In addition to the SNR effect, we find that approximately 50\% of our target clusters—predominantly in the low-mass regime—exhibit spectra that are heavily subject to stochastic sampling effects. These low-mass clusters harbor a larger number of massive stars than statistically expected, thereby artificially elevating their overall luminosities and altering their spectral profiles. Nevertheless, their underlying IMFs can be reliably constrained through our simulation framework when compared with the control runs adopting standard Salpeter/Kroupa IMFs. Furthermore, in the case of the four more massive clusters where such stochastic effects are less pronounced, the observed fluxes are more favorably explained by the underlying IMFs characterized by the $\Gamma$ values derived in this study

While the observed relationship between cluster mass and the underlying IMF across the two galaxy systems may stem from disparate star-forming environments, it is critical to emphasize that further investigation is required. Future studies should specifically examine the factors driving this relationship, including systematic uncertainties inherent in simple stellar population models and the potential presence of multiple stellar populations within the observed clusters. 

\section{Summary} \label{sec:5}
We have initiated a survey of bright young clusters in nearby galaxies to examine the diversity of the stellar IMF. This study, representing the second installment of our scientific campaign, investigates a total of 11 young clusters in the barred galaxy NGC 1313. The star-forming environment in this galaxy is characterized by subsolar metallicity, providing a useful testbed to investigate the IMF under environmental conditions distinct from those of other well-studied galaxies. Our main findings are summarized below.

The targets of this study include several 
very young clusters. Indeed, the spectra 
of four clusters exhibit strong emission 
lines. We derived the physical parameters 
(age, stellar mass, and the underlying 
IMF) of ten clusters by performing spectral 
matching with the \texttt{STARBURST99} simple 
stellar population model. 
The ages of seven clusters are younger 
than 10 Myr. With exception of ID11 (2.5 Myr), 
the youngest cluster is ID04 (5.0 Myr), 
whereas the oldest cluster is Cluster ID 9 
(300 Myr). The logarithmic ages estimated 
in this study is, on average, similar to 
those from the LEGUS survey. 

The cluster masses range from $2.8\times 10^3 M_{\sun}$ 
to $2.6 \times 10^5 M_{\sun}$. These results 
are in good agreement with those derived by the LEGUS 
survey. Half of the targets are subject to the 
effects of stochastic sampling. Nevertheless, their 
underlying IMFs were inferred through Monte Calro realizations.
Finally, the observed clusters tend to host the underlying 
IMFs with power-law index $\Gamma$ smaller than the 
standard Salpeter/Kroupa IMF. Furthermore, these $\Gamma$ values correlate with cluster masses, where more massive clusters possess underlying IMFs characterized by smaller $\Gamma$. A similar pattern has been reported for young massive clusters in the starburst galaxies NGC 4038/9. 

We have discussed the potential factors influencing this relationship between the underlying IMF and cluster mass, finding no significant evidence of systematic uncertainties in our analysis. Our results suggest that the distinct IMF-mass relationships observed in the two galactic systems are likely driven by their disparate star-forming environments. While these results 
provide key insights, further studies are essential 
to reach a definitive conclusion on the diversity 
of the stellar IMF. In particular, since this study 
relied on a single simple stellar population model 
and averaged spectra of two metallicities ($Z = 0.002$ 
and $Z = 0.014$), future work must involve rigorous 
testing against resolved stellar populations 
to validate the robustness of the model.

\begin{acknowledgments}
The authors thank the anonymous referee for constructive comments and suggestions. 
This paper has made use of data obtained under 
the K-GMT Science Program (PID: GEMINI-KR-2023B-002) 
supported by the Korea Astronomy and Space Science
Institute (KASI) grant funded by the Korean government (MSIT;
No. 2023-1-860-02, International Optical Observatory Project). 
This research has also made use of the SIMBAD database,
operated at CDS, Strasbourg, France. This work was supported by 
the National Research Foundation of Korea (NRF) grant funded by the Korean government (MSIT; grant No. RS-2022-NR072247). 
\end{acknowledgments}

%

\vspace{5mm}
\facilities{Gemini South:8.1m}


\software{Astropy \citep{2013A&A...558A..33A,2018AJ....156..123A,2022ApJ...935..167A}}





\bibliography{imf}{}
\bibliographystyle{aasjournal}



\end{document}